\documentclass[twocolumn]{emulateapj}
\pdfoutput=1
\usepackage{graphicx}
\usepackage{natbib}
\usepackage{multirow}
\usepackage{textcomp}
\usepackage{xspace}

\newcommand{\bc}{\begin{center}}
\newcommand{\ec}{\end{center}}

\newcommand{\kms}{km~s$^{-1}$}
\newcommand{\cmm}{cm$^{-3}$}
\newcommand{\cmmm}{cm$^{-3}$}
\newcommand{\msun}{M$_{\odot}$}
\newcommand{\hii}{\ion{H}{2}}

\newcommand{\am}{NH$_{3}$}
\newcommand{\cyano}{HC$_3$N}
\newcommand{\form}{H$_2$CO}

\begin{document}

\slugcomment{}

\title{The Dense Gas Fraction in Galactic Center Clouds}

\author{E.A.C. Mills}
\affil{Physics Department, Brandeis University, 415 South Street, Waltham, MA 02453}
\email{elisabeth.ac.mills@gmail.com}

\author{A. Ginsburg\altaffilmark{1}}
\affil{National Radio Astronomy Observatory\altaffilmark{3} 1003 Lopezville Rd Socorro, NM 87801}

\author{K. Immer}
\affil{Joint Institute for VLBI ERIC, Oude Hoogeveensedijk 4 7991 PD, Dwingeloo, The Netherlands}

\author{J.M. Barnes\altaffilmark{2}}
\affil{National Radio Astronomy Observatory\altaffilmark{3} 1003 Lopezville Rd Socorro, NM 87801}

\author{L. Wiesenfeld}
\affil{Universit\'{e} Grenoble Alpes, CNRS, IPAG, 38000 Grenoble, France}

\author{A. Faure}
\affil{Universit\'{e} Grenoble Alpes, CNRS, IPAG, 38000 Grenoble, France}

\author{M.R. Morris} 
\affil{Department of Physics and Astronomy, University of California, 430 Portola Plaza, Box 951547 Los Angeles, CA 90095-1547}

\author{M.A. Requena-Torres}
\affil{Department of Astronomy, University of Maryland, College Park, MD 20742, USA}

\altaffiltext{1}{A. Ginsburg is a Jansky Fellow of the National Radio Astronomy Observatory.}
\altaffiltext{2}{J.M. Barnes was a summer research student at the National Radio Astronomy Observatory.}
\altaffiltext{3}{The National Radio Astronomy Observatory is a facility of the National Science Foundation operated under cooperative agreement by Associated Universities, Inc.}

\begin{abstract}
We present an analysis of gas densities in the central R=300 parsecs of the Milky Way, focusing on three clouds: GCM-0.02-0.07 (the 50 km/s cloud), GCM-0.13-0.08 (the 20 km/s cloud), and GCM0.25+0.01 (the ``Brick"). Densities are determined using observations of the $J$=(3-2), (4-3), (5-4), (10-9), (18-17), (19-18), (21-20), and (24-23)
transitions of the molecule \cyano. We find evidence of at least two excitation regimes for \cyano\, and constrain the low-excitation component to have a density less than $10^4$ cm$^{-3}$ and the high-excitation component to have a density between $10^5$ and $10^6$ cm$^{-3}$. This is much less than densities of $10^7$ cm$^{-3}$ that are found in Sgr B2, the most actively star-forming cloud in the Galactic center. This is consistent with the requirement of a higher density threshold for star formation in the Galactic center than is typical in the Galactic disk. We are also able to constrain the column density of each component in order to determine the mass fraction of `dense' (n$>10^5$ cm$^{-3}$) gas for these clouds. We find that this is $\sim$15\% for all three clouds. Applying the results of our models to ratios of the (10-9) and (3-2) line across the entire central R=300 pc, we find that the fraction of gas with n$>10^4$ \cmmm increases inward of a radius of $\sim$140 pc, consistent with the predictions of recent models for the gas dynamics in this region. Our observations show that \cyano\, is an excellent molecule for probing the density structure of clouds in the Galactic center. 

\end{abstract}


\section{Introduction}

The inner R$\sim$300 parsecs of the Milky Way hosts a  concentration of molecular gas known as the Central Molecular Zone or `CMZ'. Slightly less than 5\% of the total molecular gas reservoir of $\sim8.4\times10^8$ \msun\, in the Galaxy is estimated to be contained in this region \citep{Dahmen98, Nakanishi06}. In the Milky Way's CMZ, the bulk of the gas at high column density is found in a ring of giant molecular clouds, orbiting the central supermassive black hole at radii of 50 to 100 pc \citep{Molinari11,Kruijssen15}, on so-called `x2' orbits \citep{Binney91}. Recent theory predicts gas accumulates at these radii due to a minimum in the shear which is responsible for its radial transport \citep{KK15}. Similarly-sized and shaped central concentrations of gas are frequently seen in the centers of barred galaxies, including nearby galaxies like IC342 \citep{Ishizuki90}, M83 \citep{Elmegreen98}, the starburst NGC 253 \citep{GB00} and the Seyfert 2 galaxy NGC 4945 \citep{Chou07}. In the CMZ, the distribution of this gas is asymmetric in several ways, with more gas found both at positive longitudes and on the near side of this ring, which is suggested to be due to some combination of instabilities in inflowing gas \citep{Sormani18}, and tidal compression at different points along the orbit \citep{Longmore13b}. 

Inside the CMZ, the properties of the gas are much more extreme than in the Galactic disk. Gas temperatures are measured to range from 50-400 K \citep{Gusten85,Huttem93b,Ao13,Mills13a,Ginsburg16,Krieger17}, significantly elevated over the dust temperatures, which range from 20-50 K \citep[][Battersby et al. in prep.]{Longmore12,Molinari11}. Linewidths due to turbulent motions are also large, with $\sigma$ ranging from 0.6-20 \kms\, over size scales of 0.2-2 pc \citep{Shetty12,Kauffmann17a,MCHH09}, though it should be noted that on scales $\gtrsim$0.5 pc, some of these large linewidths are also due to coherent velocity gradients from both orbital motion and strong shearing \citep[e.g.,][]{Federrath16}. Finally, as a result of either the increased temperature or turbulence (or both), the gas is also extremely chemically rich, with 'hot core' molecules widespread and abundant in this region \citep{RT06,Jones12,Jones13}. The elevated densities and temperatures as well as the rich chemistry seen in the CMZ are similar to conditions observed in the nuclei of nearby galaxies like NGC 253, IC 342, and Maffei 2  \citep{Paglione95,Henkel00,Ott05,Meier05,Meier12,Meier15,Gorski17,Gorski18}. 

Gas densities are also higher than those found in the disk, with the average density in the CMZ canonically taken to be $10^4$ cm$^{-3}$ \citep{Bally87,Gusten83} and values for different gas components ranging from $10^2$ cm$^{-3}$ to $10^7$ cm$^{-3}$ \citep{Walmsley86,Lis91,Dahmen98,Magnani06,Goto11,Etx11,RT12,Mills13b}. However, the various methods used to determine gas densities have key uncertainties. Density measurements based on the critical density of individual molecular transitions \citep[e.g.,][]{Bally87,Longmore13} can be significantly in error based on the actual gas excitation conditions \citep{Shirley15}. Gas densities inferred from column densities depend sensitively on an assumption of 3D shape, which can lead to errors if clouds are elongated along the line of sight \citep{Longmore12, Henshaw16a}. Excitation-derived gas densities have the potential to be more accurate than either of these methods. However, in practice these are often subject to significant uncertainties due to the degeneracy between temperature and density in driving the excitation for most molecules \citep{Dahmen98}. Moreover, nearly all estimates of density derived from excitation apart from those determined toward a few exceptional sources \citep[e.g., the CND and Sgr B2;][]{Lis91,RT12,Mills13b} have assumed there is only a single density component present in the gas \citep{Gusten83,Serabyn91,Serabyn92,Zylka92,Tanaka18}. Thus, the distribution of gas densities in the CMZ, and density structure of individual clouds are both poorly constrained. 

Measurements of cloud density are especially pertinent for understanding why the Galactic center has so little star formation \citep[$\sim5\%$ of the total star formation rate of the Milky Way as a whole;][]{Chomiuk11,YZ09,Crocker11,Longmore13,Koepferl15}. While this might appear consistent with the fraction of the Galaxy's molecular gas which lies in the CMZ (also $\sim$5\%), the large fraction of gas in this region that is 'dense' (n$>10^4$) appears to violate `laws' relating the amount of gas to the amount of star formation \citep{Longmore13,Lada13}. Determining whether the CMZ is truly a discrepant point in these universal relationships \citep[e.g., a universal threshold of gas density needed to form stars anywhere in a galaxy may not be valid in an environment like the Galactic center with much stronger shear or larger turbulence][]{Kruijssen14} first requires a careful assessment of whether we are undercounting the amount of star formation \citep[e.g.,][]{Lu17}, or overestimating the dense gas fraction. 

We present measurements of density in three CMZ clouds: GCM-0.02-0.07 and GCM-0.13-0.08 (the 50 and 20 \kms\, clouds in the Sgr A complex), and GCM 0.25+0.01 (the so-called `Brick' cloud in the dust ridge between Sgr A and Sgr B2) using multiple transitions of \cyano. In the high column-density environment of the CMZ, the relatively low abundance of \cyano\, relative to H$_2$ \citep[ $\sim3\times10^{-8}$, assuming a CO to H$_2$ abundance of $10^{-4}$;][]{Walmsley86} make it an optically-thin tracer, as opposed to molecules like CO and HCN which become optically-thick or self-absorbed \citep{Dahmen98,Mills17a}. Additionally, as a relatively heavy molecule \cyano\, has closely-spaced rotational transitions, allowing for efficient observation of transitions spanning a huge range of critical densities. Finally, \cyano\, has a simple, linear structure, with well-determined collisional coefficients calculated up to temperatures of 300 K, appropriate for the Galactic center environment \citep{Faure16}. This makes \cyano\, an ideal tracer for measuring density structure in CMZ clouds. In Section \ref{obs} we describe the setup of observations used in our analysis. The measured parameters of the \cyano\, emission and our excitation modeling are presented in Section \ref{RA}.  We conclude in Section \ref{dis} with a discussion of the derived densities, and their relevance to understanding star formation in the CMZ environment.

\section{Observations and Data Calibration}
\label{obs}

The data used in the density analysis in this paper were obtained from three sources: the 100 m Green Bank telescope (GBT) of the Green Bank Observatory\footnotemark[1]\footnotetext[1]{The Green Bank Observatory is a facility of the National Science Foundation and is operated by Associated Universities, Inc.}, the 22 m Mopra radio telescope in Australia, and the 12 m Atacama Pathfinder EXperiment (APEX) telescope in Chile.

In total, 8 rotational transitions of \cyano\, were observed: J = (3-2), (4-3), (5-4), (10-9), (18-17), (19-18), (21-20) and (24-23). \
We also observed the $^{13}$C isotopologues of \cyano\, in the J = (3-2) and (5-4) transitions. 
Properties of all observed transitions including rest frequencies are given in Table \ref{HC3N}.

\subsection{Mopra}

\begin{figure}[tbh]
\vspace{0.1cm}
\hspace{-1cm}
\includegraphics[scale=0.65]{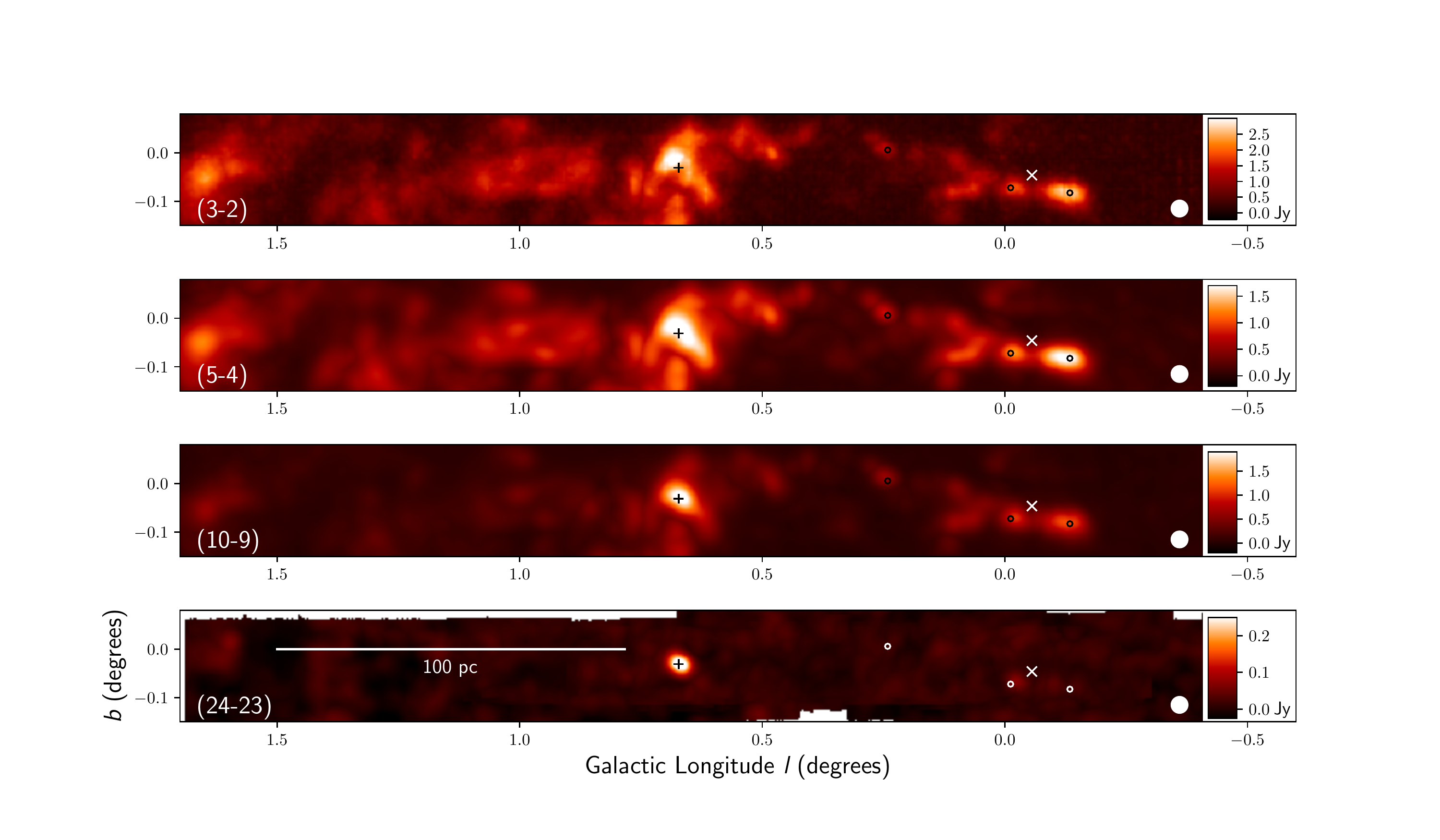}
\caption{Maps of peak emission in the CMZ for the $J$=(3-2), (5-4), (10-9), and (24-23) transitions of \cyano, smoothed to a 120$''$ beam. The size of the smoothed beam is shown as a filled white circle. The white 'x' marks the location of Sgr A*, and the black cross marks the location of Sgr B2(N). Circles indicate the positions for which there are additional, pointed observations made with the GBT in the $J$=(3-2),(4-3), and (5-4) lines.}
\vspace{-0.7cm}	
\label{Fig1_Map}
\end{figure}

Maps of \cyano\, emission are available from several surveys conducted with the Mopra radio telescope. \cyano\, (10-9) and (5-4) line maps were taken from CMZ surveys  published in \cite{Jones12} and \cite{Jones13}. These lines were observed over a 2.5 × 0.5 deg$^2$ region covering the inner 350 parsecs of the CMZ \citep[assuming a Galactocentric distance of 8 kpc;][]{Boehle16} and including all three of the clouds studied here. \cyano\, (3-2) line maps were taken from the HOPS southern Galactic plane survey \citep{Walsh11}. The description of the data calibration is given in the survey papers. Calibrated and imaged survey data were obtained from the Australia Telescope Online Archive\footnotemark[2]\footnotetext[2]{https://atoa.atnf.csiro.au/}. The (10-9) data have a spatial resolution of 40$''$, a velocity resolution of $\sim$2 \kms, and a per-channel RMS noise of 50 mK. The (5-4) data have a spatial resolution of 65$''$, a velocity resolution of $\sim$1.8 \kms, and a per-channel RMS noise of 37 mK. The (3-2) data have a spatial resolution of 120$''$, a velocity resolution of $\sim$0.4 \kms, and a per-channel RMS noise of 200 mK. Integrated emission maps of all these lines in the CMZ are shown in Figure \ref{Fig1_Map}. The estimated uncertainty of the amplitude calibration of these data is $<25$\%. 

\subsection{GBT}

We have also made pointed observations of the \cyano\, (3-2), (4-3) and (5-4) transitions with the GBT as part of a survey for highly-excited \am\, emission published in \cite{Mills13a}. The data were observed with the dual-beam Ka-band and Q receivers between May 12 and May 29, 2009. We employed a position-switching technique, using an offset position of Right Ascension, Declination) = (17$^{\mathrm{h}}$46$^{\mathrm{m}}$00$^{\mathrm{s}}$,  -$28\degr13'57''$), for clouds east of ($l=0.1\degr$), and an offset position of (Right Ascension, Declination) = (17$^{\mathrm{h}}$45$^{\mathrm{m}}$59.9$^{\mathrm{s}}$, -29$\degr 16' 47''$) for more westerly clouds. With the GBTIDL\footnotemark[3]\footnotetext[3]{http://gbtidl.nrao.edu/} reduction and analysis software, we corrected the antenna temperature of the observed targets for the frequency-dependent opacity at the observed elevation. We also used observations of the flux calibrator 3C286 to more accurately determine the relative amplitude calibration of the data, which is otherwise limited to $10-15\%$ accuracy by the temporal fluctuations of the noise diode. We estimate the uncertainty of our relative amplitude calibration to be $<10\%$ based on the RMS fluctuations in the observed spectrum of the phase calibrator. Finally, we assumed that the observed emission is extended over an area larger than the telescope beam, and applied a frequency-independent correction of 1.32 for the main-beam efficiency. Additional details pertaining to the observations and calibration of these data are given in \cite{Mills13a}. 

Observations of all lines observed with the GBT were made in a single, pointed beam toward each cloud. For GCM0.25+0.01, the pointing was centered on (R.A., Dec.) = (17$^{\mathrm{h}}$46$^{\mathrm{m}}$10.3$^{\mathrm{s}}$,  -28\degr 43\arcmin 37.0\arcsec). For GCM-0.02-0.07, the pointing was centered on (R.A., Dec.) = (17$^{\mathrm{h}}$45$^{\mathrm{m}}$52.4$^{\mathrm{s}}$, -28\degr 59\arcmin 02.0\arcsec).  For GCM-0.13-0.08, the pointing was centered on  (R.A., Dec.) = (17$^{\mathrm{h}}$45$^{\mathrm{m}}$37.9$^{\mathrm{s}}$, -29\degr 03\arcmin 52.0\arcsec).
The spatial resolution of the data ranged from 17$''$ for the (5-4) line to 28'' for the (3-2) line. The spectral resolution of the observations was 390.625 kHz, or 3.1 to 4.4 \kms over the observed range of frequencies, sufficient to resolve lines with intrinsic widths of 15 to 30 \kms. The positions of these pointings are shown in Figure \ref{Fig1_Map}.

Finally, for the GCM-0.02-0.07 cloud, a 4$'$ by 4$'$ map centered on  (R.A., Dec.) = (17$^{\mathrm{h}}$45$^{\mathrm{m}}$52$^{\mathrm{s}}$,  -28\degr 59\arcmin 03\arcsec) was also observed using the GBT in the \cyano\, $J$ = (5-4) line. 

\subsection{APEX}

Observations of the \cyano\, (24-23) line were made with the 230 GHz facility receiver. These observations comprised a 120$'\times16'$ map of the entire 300 pc inner region of the CMZ as part of the observations presented in \cite{Ginsburg16}. Details of these observations and the procedures for calibration and imaging are given in this paper. The spatial resolution of these data is $\sim29''$ and the velocity resolution is 1 \kms. The per-channel RMS noise near the \cyano\, (24-23) line is $\sim$0.03 K.  The estimated uncertainty of the amplitude calibration of these data is $\sim$15\%. 

Observations of the \cyano\, (18-17), (19-18) and (21-20) lines were also made with the SEPIA receiver \citep{Belitsky18} covering small mosaics toward the three clouds studied here. The spatial resolution of these data is $\sim38''$ and the velocity resolution is 1 \kms. The per-channel RMS noise near the observed \cyano\, lines is 0.01-0.02 K.  
The estimated uncertainty of the amplitude calibration of these data is $\sim$10\%. 

\subsection{Data Combination}
The data set studied in this paper consists of a combination of maps and pointed observations toward GCM0.25+0.01, GCM-0.02-0.07, and GCM-0.13-0.08. We conduct two separate analyses. The first is a study of the emission from all 8 \cyano\, transitions in a single beam toward the three clouds. For this analysis, we do not use maps of the (3-2) or (5-4) line, and smooth all other maps to the lowest common spatial resolution (40$''$), set by the $J$ = (10-9) observations. For these map data: the $J$ = (10-9) line observed with Mopra, and the $J$ = (18-17), (19-18), (21-20), and (24-23) lines observed with APEX, spectra are extracted for each cloud at the position of the GBT observations. As the GBT data consist of single pointings toward these positions, the resolution of the (5-4) and (4-3) line data cannot generally be perfectly matched to that of the other data sets. As a result, we estimate a larger uncertainty for these measurements. However, for the GCM-0.02-0.07 cloud, which was mapped with the GBT in the $J$ = (5-4) line, a spectrum for this line can be extracted from a 40$''$ aperture to match the resolution of the lines extracted from the other maps. 

For the second analysis, we study the ratio of \cyano\, lines over the entire CMZ using maps of the (3-2), (5-4) and (10-9) lines. For this analysis, all maps are smoothed to the resolution of the (3-2) line ($120''$). 

\section{Results and Analysis}
\label{RA}

\begin{figure}[tbh]
\includegraphics[scale=0.5]{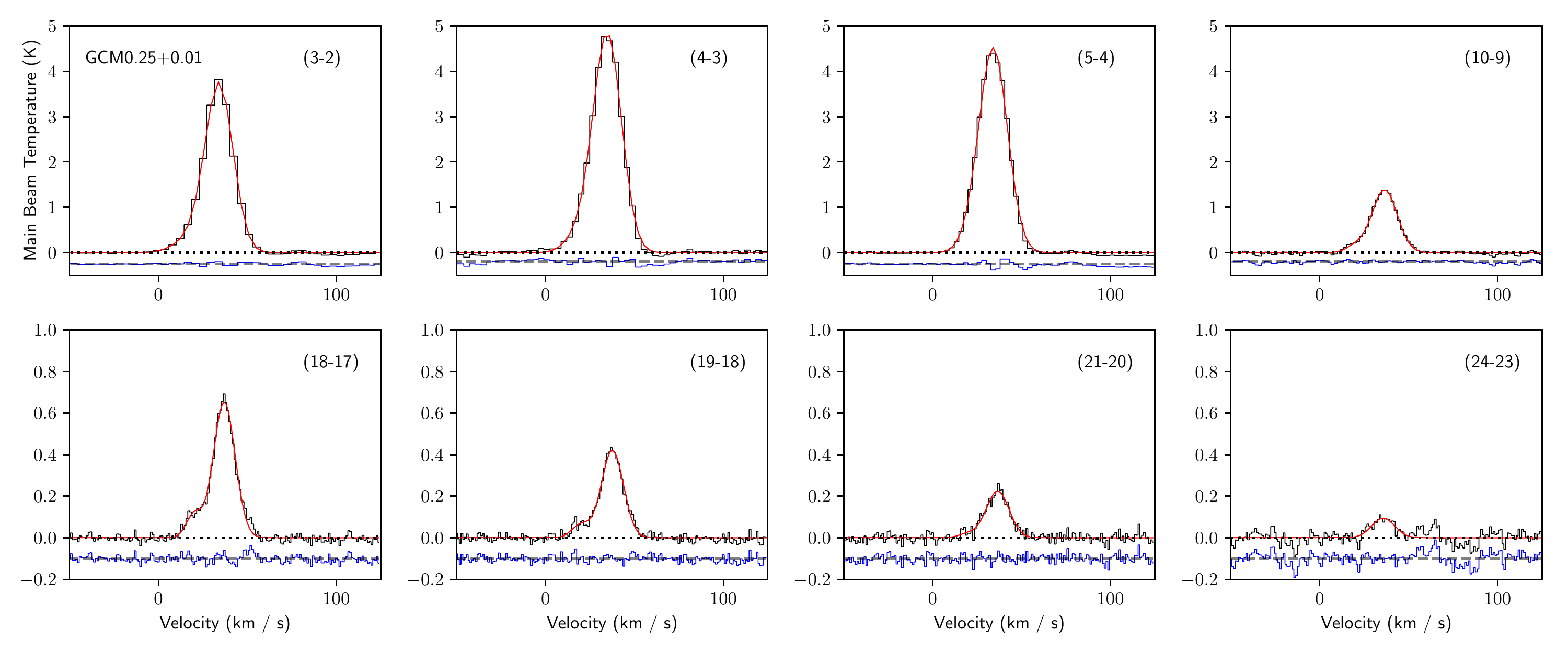}\\
\includegraphics[scale=0.5]{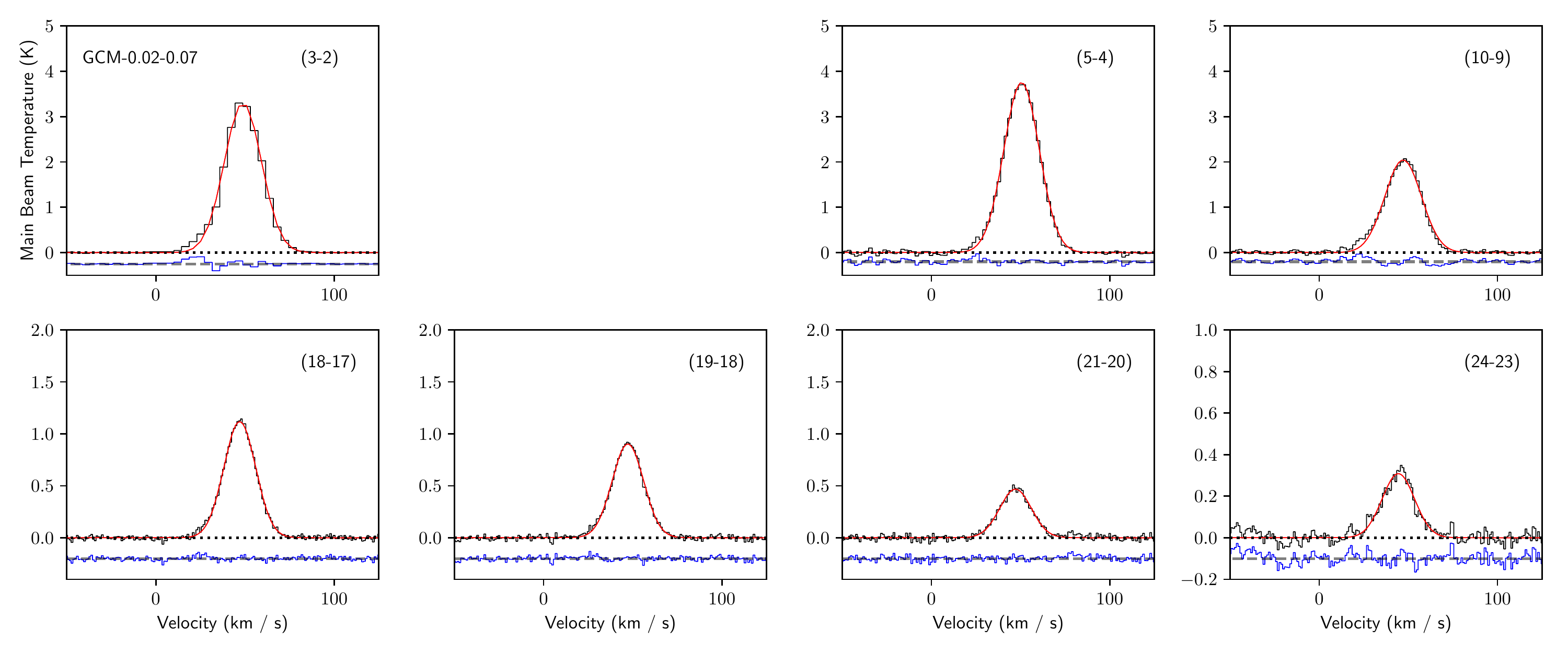}\\
\includegraphics[scale=0.5]{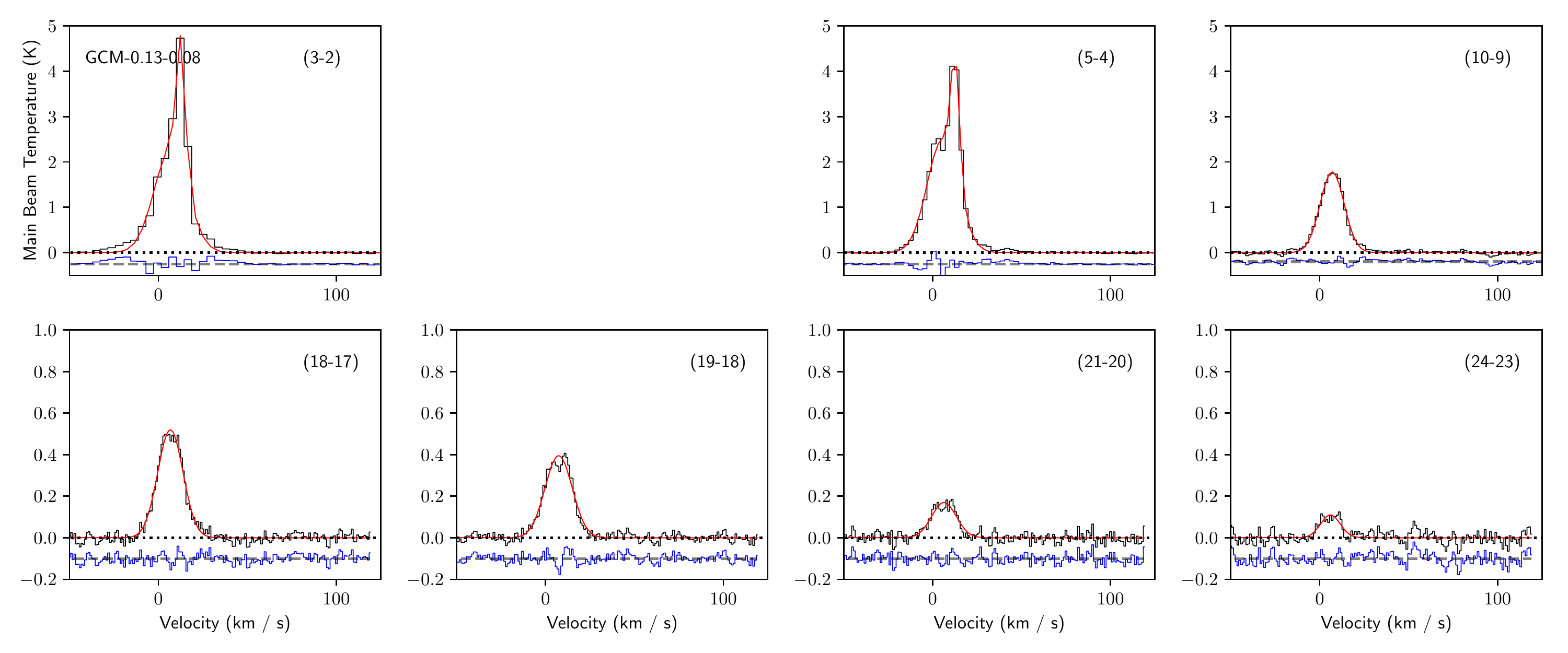}
\caption{Spectra of \cyano\, in the GCM0.25+0.01 cloud (top), the GCM-0.02-0.07 cloud (middle), and the GCM-0.13-0.08 cloud (bottom). Gaussian fits to each line are overplotted in red, and the residuals of the fits are plotted below the spectrum in blue.}
\label{Fig2}
\end{figure}

\begin{figure}[tbh]
\includegraphics[scale=0.5]{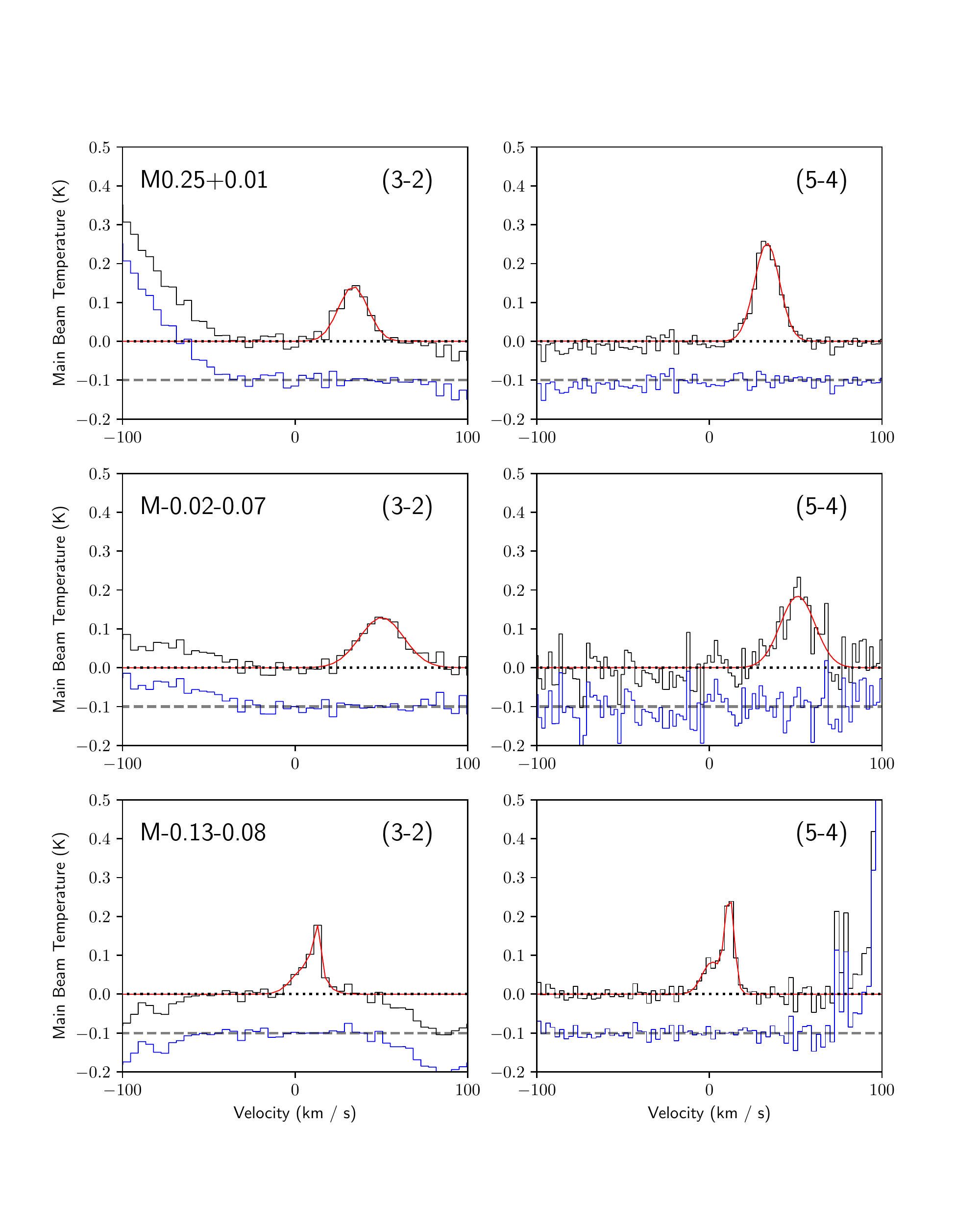}
\caption{Spectra of H$^{13}$CCCN in the GCM0.25+0.01 cloud (top), the GCM-0.02-0.07 cloud (middle), and the GCM-0.13-0.08 cloud (bottom).}
\label{Fig3}
\end{figure}

The spectra of all of the \cyano\, lines observed toward each source are displayed in Figures \ref{Fig2} and \ref{Fig3}. Toward GCM0.25+0.01, all eight main transitions of the main isotopologue of \cyano\, are observed and detected, as well as the two transitions of H$^{13}$CCCN. Toward GCM-0.02-0.07 and GCM-0.13-0.08, we do not have observations of the \cyano\, (4-3) line, but all other lines are observed and detected. For all detected transitions, we fit the spectral lines with Gaussian profiles using PySpecKit \citep{Ginsburg11}. 

The spectra toward GCM-0.02-0.07 show a single velocity component at $\sim$ 48 \kms, which we are able to fit with a single Gaussian profile having a width of $\sim$ 22 \kms. However, the spectra toward GCM-0.13-0.08 and GCM0.25+0.01 are somewhat more complicated, with each of these showing two discernable velocity components in many of the observed spectra. For these two clouds, we then fit up to two Gaussian profiles to their line spectra. For GCM-0.13-0.08, there is a primary velocity component with FWHM $\sim18$ \kms\, centered at $\sim$7 \kms, and a second, narrower velocity component (FWHM $\sim$6 \kms) at $\sim$12 \kms\, which is prominent in lines with $J\leq$5. We therefore fit both of these components in the $J$ = (3-2), (4-3), and (5-4) lines of \cyano\, and H$^{13}$CCCN, but only fit a single component in the higher-$J$ lines. For GCM0.25+0.01, there is a weak, low-velocity component at $\sim20$ \kms\, and a stronger, main line component at $\sim$35 \kms. We fix the velocity of the low-velocity component to be 20 \kms, and include it in fits of all lines except the weak $J$ = (24-23) transition. The measured line properties of each transition of \cyano\, and H$^{13}$CCCN from these fits are given in Table \ref{line_params}. 

By comparing the \cyano\, and H$^{13}$CCCN line brightness temperatures, we can estimate the optical depth of the $J$ = (3-2) and (5-4) lines, by assuming the [$^{12}$C/$^{13}$C] ratio in the Galactic center to be 25 \citep{WR94,Riq10}. For each cloud, we then measure the ratio of the fluxes of the two transitions of \cyano\, and H$^{13}$CCCN, averaging together the values for multiple Gaussian components, if present. Using the $J$=3-2 ($J$=5-4) transitions, we subsequently find \cyano\, to H$^{13}$CCCN line ratios of 26$\pm$2 (22$\pm$2)  for GCM0.25+0.01, 21$\pm1$ (20$\pm$1) for GCM-0.02-0.07, and 33$\pm$6 (25$\pm$7) for GCM-0.13-0.08. In all cases, the ratio derived from the  $J$ = (5-4) line is slightly smaller than that from the $J$ = (3-2) line, consistent with the (5-4) line being somewhat less optically thin \citep[as it is closer to the peak of the spectral line energy distribution;][]{Walmsley86}. Overall however, all of these values are consistent with the main \cyano\, transitions being optically-thin, with $\tau \lesssim$0.5. While this may not be a good assumption in the highest-column density source Sgr B2, this assumption should largely hold for the rest of the CMZ. 

\subsection{Ratio Maps}
\label{ratio}
\begin{figure*}
\hspace{-1cm}
    \includegraphics[scale=0.65]{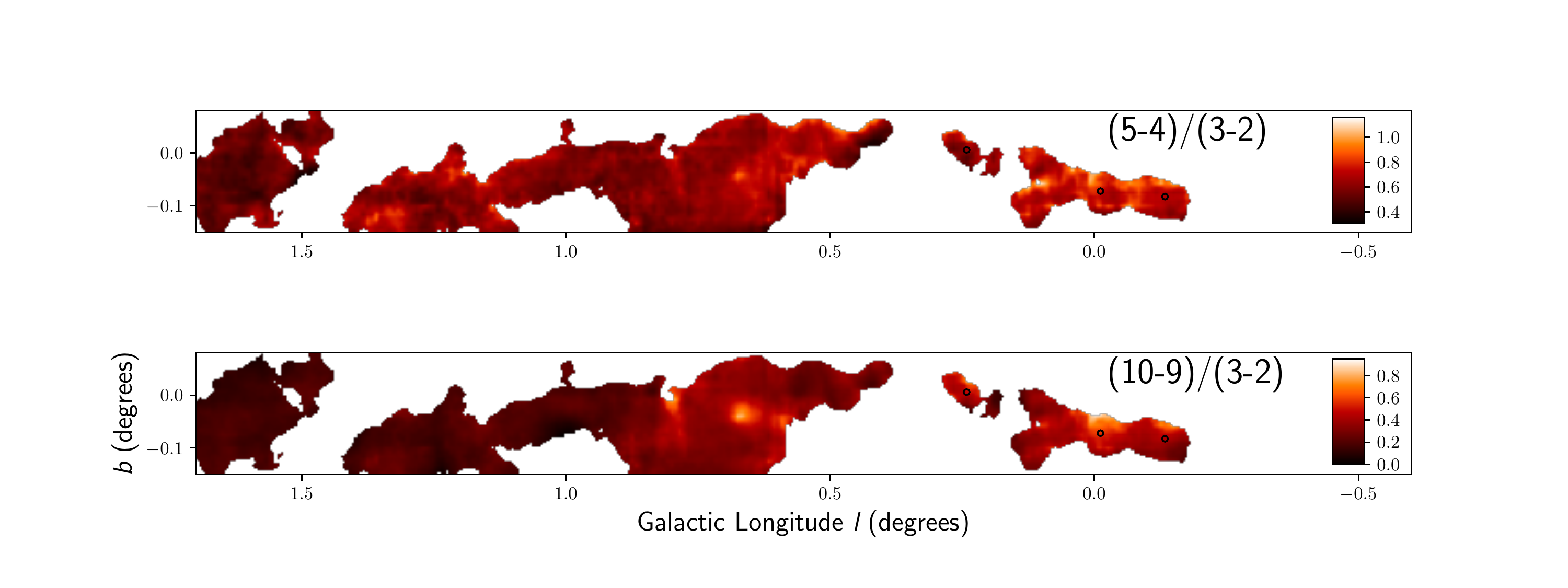}
    \caption{Ratio maps of the \cyano\, $J$= (5-4) to (3-2) lines (top) and $J$= (10-9) to (3-2) lines (bottom).  As in Figure \ref{Fig1_Map}, circles indicate the positions of pointed observations made with the GBT. A qualitative difference between the inner CMZ (right half)
    and the 100$<$r$<$200 pc region (left half) is evident: the excitation is lower beyond the orbit of Sgr B2.
    This difference suggests that the gas in the outer CMZ has a lower average density and/or lacks
    a high-density (high-excitation) component, consistent with models in which the CMZ is fed by
    highly turbulent, non-star-forming gas driven in by the Galactic bar.}
    \label{Fig4}
\end{figure*}

Using the Mopra maps of the (3-2), (5-4) and (10-9) lines of \cyano\, we then construct maps of ratios between these lines across the entire CMZ (the 24-23 map is not used as this transition is only detected in a few isolated positions beyond the three clouds already probed). The resulting ratio maps are shown in Figure \ref{Fig4}. 

Before attempting to identify trends, it should be noted that the Sgr B2 core at $l\sim0.4\degr$ exhibits unresolved absorption against its numerous \hii\, regions in the $J$ = (3-2) and to a slightly lesser extent in the $J$ = (5-4) line, causing an apparent peak in the ratio maps which is not due to differing excitation in this source. Ignoring this area, we see that the (5-4) / (3-2) line ratio does not change greatly over the CMZ. The largest median (5-4) / (3-2) ratios, 0.74$\pm$0.09, are found in the Sgr A complex between $l$ = -0.2$\degr$ and $l$ = +0.2$\degr$ and including the GCM-0.02-0.07, and GCM-0.13-0.08 clouds. The ratio is slightly smaller (0.61$\pm$0.09) in the gas between l = +0.2 and l = +0.4, which includes GCM0.25+0.01, the 'dust ridge' \citep{Lis94,Immer12}, and the gas around Sgr B2. Finally, the smallest ratio, 0.52$\pm$0.06, is found in the gas at $l > 1.5\degr$. This trend is slightly modified when we examine the map of the ratio of the $J$ = (10-9)  to $J$ = (3-2) lines, which are more widely spaced in excitation energy. Here, gas in the outer CMZ at $l>1\degr$ has a smaller median (10-9)/(3-2) ratio of 0.18$\pm$0.04, compared to a median ratio of 0.33$\pm$0.11 in gas with |$l$| $< 1\degr$.  

As we find the brightest lines, $J$ = (3-2) and (5-4) are optically thin in clouds representative of the gas outside of the Sgr B2 core (and the weaker, higher $J$ lines should follow suit), these trends should not be driven by differences in line opacity. Rather, they are indicative of changes in temperature, density, and/or the relative abundance of low-excitation and high-excitation \cyano\. In particular, the observed distribution of the \cyano\, (10-9)/(3-2) ratio shows that there is more high-excitation gas in the inner R = 1$\degr$ / 140 parsecs. As a systematic temperature difference is not seen between the inner and outer CMZ gas \citep{Ginsburg16,Krieger17}, this is more likely to be a density effect.The inner CMZ gas would then have a higher average gas density, either because clouds in the inner CMZ have a larger fraction of gas in a high-density gas component than those in the outer CMZ, or because the gas in the inner CMZ attains higher peak densities than that in the outer CMZ. We discuss the possibility that this result could also be affected by the relative abundances in Section \ref{abund}. 

To better understand the quantitative constraints on the physical conditions of the gas from these maps, we next turn to non-LTE modeling of the \cyano\, lines. Given the relatively low $J$-numbers of the mapped transitions, excitation modeling of just these three lines is unable to robustly constrain the full physical conditions of the gas in this region. However, as we have three clouds for which we have observed additional high-$J$ \cyano\, lines, we attempt to gain some insight into the global gas densities in the CMZ by modeling \cyano\, excitation in these clouds.

\subsection{Excitation Modeling}

We use the publicly-available statistical equilibrium radiative transfer code RADEX \citep{vanderTak07}, a zero-dimensional non-LTE code employing escape probability formalism to model the observed line intensities as a function of the physical conditions. The escape probability method simplifies the radiative transfer calculation by assuming that photons either completely escape the source (the likelihood of this is dependent on the local opacity), or are immediately absorbed at the same location where they were emitted. The radiative coefficients for the observed rotational transitions of \cyano\, are taken from the Cologne Database for Molecular Spectroscopy \citep[CDMS;][]{Muller05}, and the collisional coefficients are taken from \cite{Faure16}. Two collisional partners, ortho-H$_2$ and para-H$_2$ are considered, and we adopt a statistical ortho-to-para ratio for H$_2$ (3:1), however varying this does not appear to have a strong effect on the results. 

RADEX takes as input parameters the kinetic temperature, the number density of H$_2$, the column density of \cyano\, per velocity element, and the line full width at half maximum (these last two values are used by RADEX to determine the local opacity). Additionally, a background radiation temperature is specified; this value is held fixed for all of the fitting we conduct. We adopt a representative FWHM value for each individual cloud: 16 \kms\, for GCM0.25+0.01, 22 \kms\, for GCM-0.02-0.07, and 28 \kms\, for GCM-0.13-0.08. We fit a two-component model of temperature and density for all three clouds. We consider kinetic temperatures between 20 and 300 K for both excitation components. For the low-excitation component, we consider H$_2$ number densities between $10^2$ cm$^{-3}$ and $10^5$ cm$^{-3}$, and \cyano\, column densities between $10^{14}$ cm$^{-2}$ and $10^{16}$ cm$^{-2}$. For the high-excitation component, we consider H$_2$ number densities between $10^4$ cm$^{-3}$ and $10^7$ cm$^{-3}$, and \cyano\, column densities between $10^{13}$ cm$^{-2}$ and $10^{15}$ cm$^{-2}$. 

Using RADEX, we then compute line intensities for the three fixed FWHM values and a range of temperatures, densities, and column densities. We evaluate the line intensities over a 30 by 30 grid of input temperatures (linearly spaced) and densities (logarithmically spaced), with a separate grid computed for each of 20 logarithmically-spaced values of the column density. This procedure is repeated for both the low-excitation and high-excitation components, to make two separate sets of output line intensities. In addition to reporting the brightness temperature and flux for all transitions of \cyano\, between $J$ = (1-0) and $J$ = (24-23), RADEX also reports the line opacity, which we use to verify that the modeled (3-2) and (5-4) lines are optically-thin, consistent with our observations. 

\begin{figure}[tbh]
\includegraphics[scale=0.35]{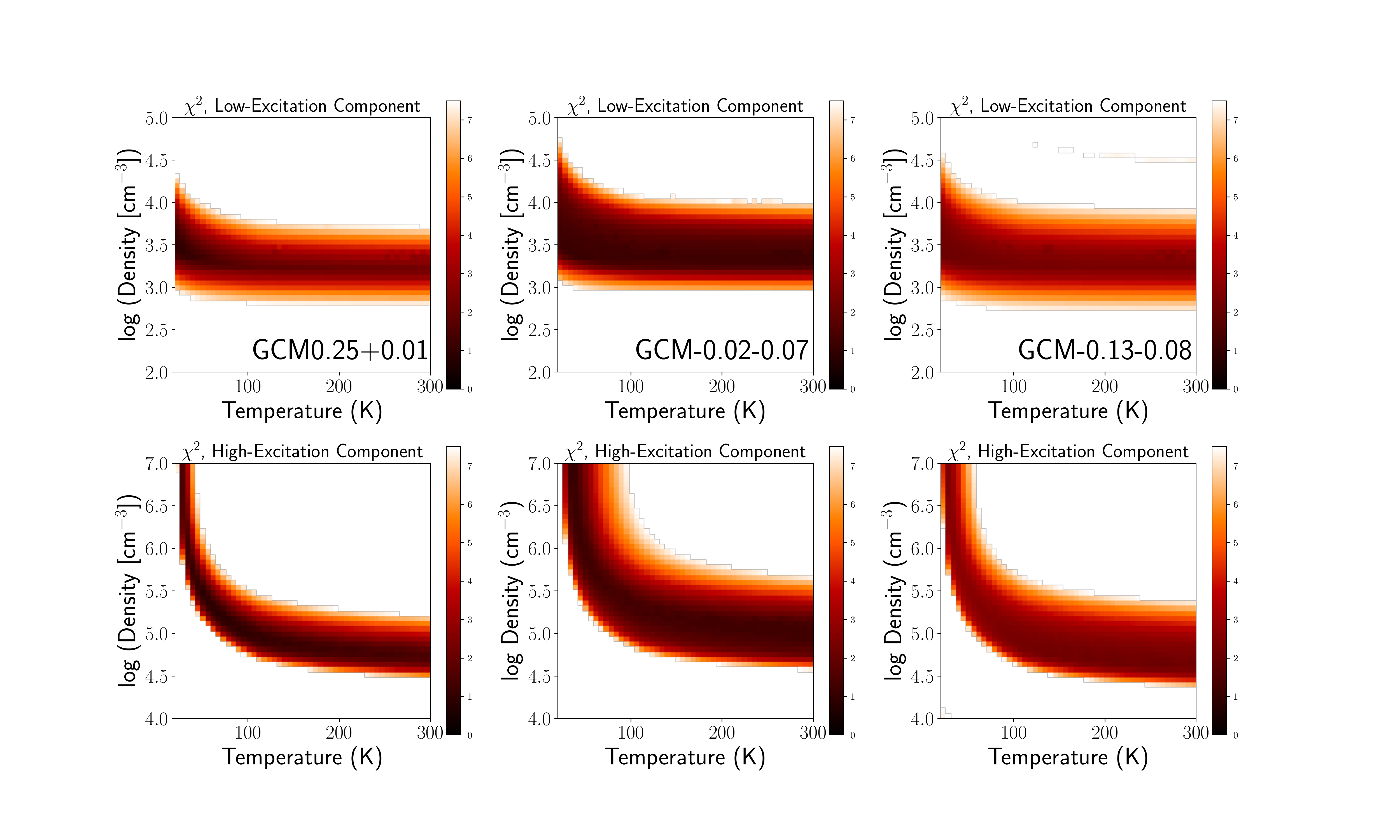}\\
\caption{Chi-squared values from fitting computed grids of density and temperature values from a two-component RADEX fit to the flux of the observed \cyano\, lines. Fits are shown separately for GCM0.25+0.01 {\bf (Left)}, GCM-0.02-0.07 {\bf (Center)}, and GCM-0.13-0.08 {\bf (Right)}. {\bf Top Row:} Temperature and density constraints on the low-excitation component  {\bf Bottom Row:} Temperature and density constraints on the high-excitation component.   }
\label{Fig5}
\end{figure}

\begin{figure}[tbh]
\includegraphics[scale=0.35]{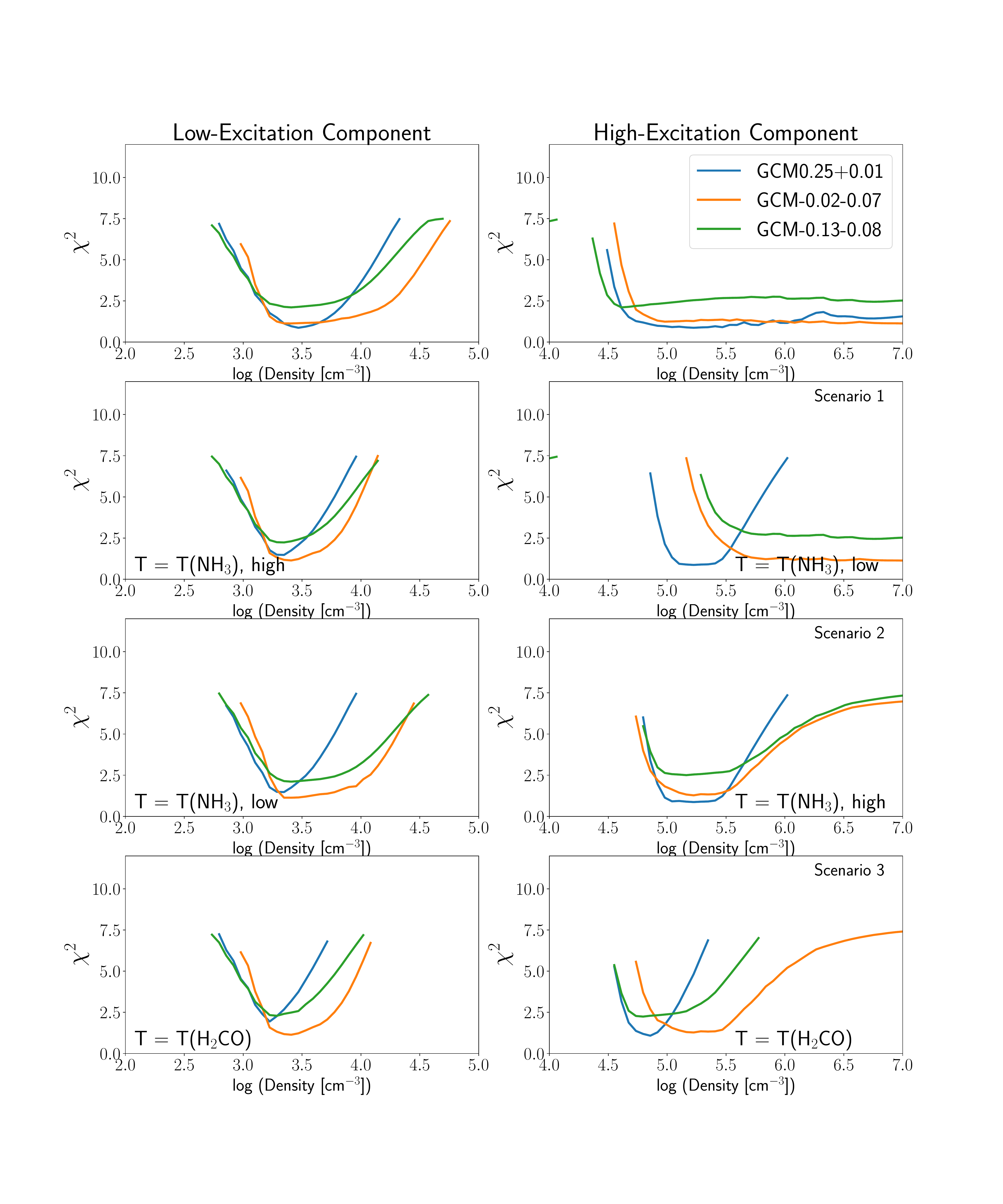}
\caption{Chi-squared values from RADEX fitting for the density of the low- and high- excitation components for each of the three observed clouds: GCM 0.25+0.01, GCM-0.02-0.07, GCM-0.13-0.08, marginalized over all other parameters. {\bf Top:} No constraint on the kinetic temperature is applied. {\bf Middle-Top} Scenario 1 from Section \ref{dis}: The temperature of the low- and high-density component is restricted to be within $\sim$20\% of the low- and high- kinetic temperature component, respectively, as measured from NH$_3$ lines at the source position by \cite{Krieger17}.  {\bf Middle-Bottom} Scenario 2 from Section \ref{dis}: The temperature of the low- and high-density component is restricted to be within $\sim$20\% of the high- and low- kinetic temperature component, respectively, as measured from NH$_3$ lines at the source position by \cite{Krieger17}. {\bf Bottom} Scenario 3 from Section \ref{dis}: The temperature of both components is restricted to be within $\sim$20\% of the kinetic temperature measured from H$_2$CO lines at the source position by \cite{Ginsburg16}.  }
\label{Fig6}
\end{figure}

\begin{figure}[tbh]
\includegraphics[scale=0.35]{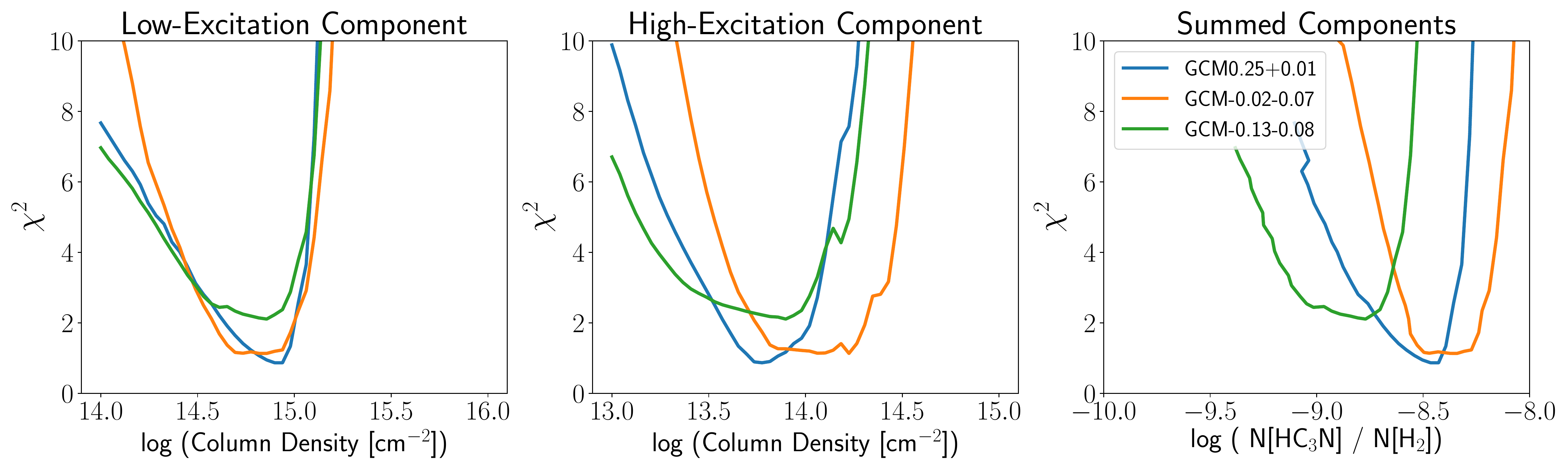}
\caption{Chi-squared values for the column density of the low- and high- excitation components ({\bf Left} and {\bf Center}) for each of the three observed clouds: GCM 0.25+0.01, GCM-0.02-0.07, GCM-0.13-0.08, marginalized over all other parameters. On the {\bf Right}, we show the best-fit abundance from summing the best-fit column densities and comparing to the average Herschel-derived H$_2$ column densities extracted from the same apertures as the \cyano\, spectra.}
\label{Fig7}
\end{figure}

To conduct a two-component fit, we then perform chi-squared fitting of the observed line fluxes (the sum of the one or two Gaussian components fit to each line) to the summed model fluxes from each combination of values in the low-excitation and high-excitation RADEX runs. The results of this fitting are displayed in Figures \ref{Fig5}, \ref{Fig6}, and \ref{Fig7}. Figure \ref{Fig5} shows the chi-squared values for fitting to each cloud as two-dimensional grids of temperature and density, marginalized over all values of the column density. Separate grids are shown for the low- and high-excitation components. Figure \ref{Fig6} shows chi-squared values from fits to each cloud as one-dimensional plots for the density, marginalized over all values of temperature and column density. The values for the low- and high-excitation components are shown separately. Figure \ref{Fig7} shows the chi-squared values from fits to each cloud as one-dimensional plots for the column density, marginalized over all values of temperature and density. Again, the values for the low- and high-excitation components are shown separately. 

In combination with the chi-squared fitting we have applied an additional joint constraint on the abundance and path length of the low-excitation component. If the gas traced by low-excitation lines of \cyano\, is truly tenuous, it could have a path length up to the entire depth of the CMZ \citep[e.g., $\sim $100 pc,][]{Kruijssen15}. However, in this case, it should trace gas with a wide range of velocities \citep[e.g., Figure 4 of ][in which velocities along a single line of sight should span $\sim$100 \kms]{Kruijssen15}, and so the relatively narrow linewidths of 15-20 \kms\, that are observed should rule this out. We adopt a maximum path length of 10 pc, which is a conservative assumption that would still mean the clouds are more extended along line of sight than they appear in the plane of the sky. 

To determine a reasonable maximum \cyano\, abundance for the low-excitation gas in the CMZ, we first determine the mean \cyano\, abundance for our best-fit model parameters. We compare the total \cyano\, column density from both the low- and high-excitation gas to an estimate of the total H$_2$ column in each cloud from Herschel-HiGal observations of the Galactic center \citep[][Battersby et al., in prep.]{Molinari11,Battersby11}. Note that is not possible to separately estimate the abundances of the individual excitation components, as the fraction of the H$_2$ column associated with each component is unknown. The resulting best-fit mean abundances are shown in the right-hand panel of Figure \ref{Fig7}, and range from 2-$5\times10^{-9}$, consistent with prior estimates for Galactic center clouds ranging between $10^{-11}-5\times10^{-9}$ for Galactic center clouds \citep{Morris76,deVicente00}. While \cite{deVicente00} suggest that \cyano\, abundances in several of the Sgr B2 hot cores could be as high as $10^{-7}$, for the extended clouds studied here we adopt a maximum value for the \cyano\, abundance in the low-excitation component of 10$^{-8}$, consistent with the most extreme values measured in the nuclei of other galaxies \citep{Mauersberger90,Aalt07b, Meier}. We can then define a column density threshold as a function of the gas volume density: 

\begin{equation}
N_{max} =  l * n *  N [\mathrm{HC}_3\mathrm{N}] / N [\mathrm{H}_2]
\label{Eq1}
\end{equation}

where $l$ is the maximum path length in cm, $N_{max}$ is the maximum allowed column density, $n$ is the modeled volume density, and $N [\mathrm{HC}_3\mathrm{N}] / N [\mathrm{H}_2]$ is the maximum allowed \cyano\, abundance. 
Eliminating those RADEX solutions that would require unphysically large path lengths or abundances sets both the lower bound on the gas densities of the low-excitation component shown in Figure \ref{Fig5} as well as the upper bound on the column densities for the low-excitation component shown in Figure \ref{Fig7}.

Overall, as expected, the temperature of these three clouds is not constrained by our fits to the \cyano\, line intensities. The density of the low-excitation gas is constrained to be between $10^3$ and $10^4$ \cmmm\, with no significant difference in this value between the three clouds. The density of the high-excitation gas is constrained to be larger than $3\times10^4$ \cmmm\, for all three clouds. If we assume the gas is warmer than 50 K as several previous studies have indicated, then the upper bound on this density is between $3\times10^5$ and $10^6$ \cmmm; without a constraint on the temperature densities up to $10^7$ \cmmm\, are allowed. The column density of the low-excitation gas is constrained to be between $\sim3\times10^{14}$ to $10^{15}$ \cmm\, for all three clouds. That of the high-excitation gas is constrained to be between $\sim3\times10^{13}$ and $3\times10^{14}$ \cmm, with somewhat higher values favored for GCM-0.02-0.07 than for the other two clouds. 
We find that the density of the low-excitation component is $<$ 10$^{4.5}$ \cmmm\, for all temperatures considered. The density of the high-excitation component is $>$ 10$^{4.5}$ \cmmm, again for all clouds and temperatures considered. 

Comparing the best-fit column densities of the low- and high-excitation components, we find that the fraction of the \cyano\, column coming from gas with $n>10^4$ \cmmm\, is $\sim$ 15\% in all three clouds. From the range of allowed column densities for each source, there is a suggestion that this fraction may be slightly lower in GCM0.25+0.01 than GCM-0.13-0.08 and GCM-0.02-0.07, however, the uncertainties are too large to make a convincing claim.

\section{Discussion}
\label{dis}

\subsection{The \cyano\, Abundance and its Implications}
\label{abund}
In the three individual Galactic center clouds studied here, we infer \cyano\, abundances ranging from 2-$5\times10^{-9}$ (a factor of 2.5 variation between the three clouds). This is comparable to \cyano\, abundances measured in the nuclei of NGC 253 \citep[$4\times10^{-9}$;][]{Mauersberger90} and IC 342 \citep[$6\times10^{-10}-3\times10^{-9}$;][]{Meier}. It is also similar to the \cyano\, abundance in the Orion hot core \citep[$2\times10^{-9}$;][]{Blake87}. This is not unexpected given the chemical similarity of the bulk CMZ gas to hot cores elsewhere in the Galaxy \citep{RT06}. However, because the abundances we measure are referenced to H$_2$ column densities from Herschel dust continuum observations, they apply only to the sum of the two observed excitation components. That we do not constrain the relative abundance of the two components is a source of uncertainty in our analysis, and for interpreting results on the dense gas fraction in the three main clouds and overall in the CMZ. 

Chemical models predict a certain amount of abundance variation for \cyano. Current chemical models for molecular clouds favor the gas-phase formation of \cyano\, via a neutral-neutral reaction \citep{Fukuzawa98,Araki16,HilyBlant18}:

\begin{equation}
\mathrm{CN} + \mathrm{C}_2\mathrm{H}_2 \rightarrow \mathrm{HC}_3\mathrm{N} + \mathrm{H} 
\end{equation}

While many chemical models of \cyano\, consider its formation in dark clouds \citep[e.g., to reproduce the abundances of the cyanopolyyne peak in the Taurus Molecular Cloud; ][]{Suzuki92}, where cold core chemistry efficiently produces both the parent species C$_2$H$_2$ as well as \cyano\, such conditions do not match what is seen in the CMZ. However, \cyano\, is also produced easily in a hot core environment like that seen globally in the CMZ, provided that its parent species C$_2$H$_2$ is liberated from grains \citep[as has been observed in hot cores;][]{Lahuis00} since C$_2$H$_2$ is not efficiently produced in the gas phase under these conditions \citep{Brown88,Charnley92}. 

In a time-dependent chemical model of a hot core at a fixed temperature and density, there is a nearly 3 order of magnitude difference between the peak \cyano\, abundance (reached in a a few $10^4$ years) and the steady state abundance reached in $\sim10^6$ years \citep{Chapman09}. As the formation reaction for \cyano\, has a weak temperature dependence, variations in temperature are expected to have only a small impact on its abundance. This is consistent with what is seen in the \cite{Chapman09} models, where changing the temperature from 200 to 100 K results in column density changes of less than 0.5 dex, while increasing the density by an order of magnitude correspondingly increases the \cyano\, column density by an order of magnitude. 

\subsubsection{The Absolute Dense Gas Fraction in individual clouds}

If we assume that the fractional \cyano\, abundance is the same in the two excitation components that we observe, then we can interpret the fraction of the \cyano\, column coming from gas with $n>10^4$ \cmmm\, ($\sim$15\%) as the actual fraction of dense molecular gas in these clouds. However, we cannot rule out that these components actually have different abundances, as we do not have a direct constraint on the abundances of the individual components.  If we allow the relative abundances of the two components to vary by an amount equal to the variation in the total \cyano\, abundance seen between the different clouds in this analysis (a factor of 2.5), then the typical dense gas fraction would be loosely constrained to be between 6\% and 37\%.  Interestingly, \cite{Morris76} in a two-component model of \cyano\, in Sgr B2 found that the denser `core' component had a lower \cyano\, abundance than the lower-density `halo' component, which if it applied more globally would result in a larger value for the dense gas fraction. A more precise estimate will ultimately require comparison of the \cyano\, results with those of CO or another proxy gas tracer, in order to infer the abundance relative to H$_2$ for each component.

\subsubsection{The Relative Dense Gas Fraction across the CMZ}
We see a difference in the ratio of the \cyano\, (10-9) line (which primarily traces the high-excitation \cyano\, component) and (3-2) line (which primarily traces the low-excitation \cyano\, component) in the CMZ for gas between $|l| < 1\degr$ and gas at $l>1\degr$. If the relative \cyano\, abundance of these two excitation components (whether or not they are identical) does not change, then we argue in Section \ref{ratio} that the different (10-9)/(3-2) line ratios are likely to be due to a difference in the gas density of the two components. However, if these components represent physically and chemically distinct gas, it is also possible that a change in the relative \cyano\, abundances in the low- and high-excitation components could reproduce this line-ratio signature. These abundance differences would be unlikely to be driven by temperature; as has been previously noted, the inner CMZ gas is not systematically warmer (or colder) than the gas at $l>1\degr$ \citep{Ginsburg16,Krieger17}. Similarly, Figure 2 of \cite{Mills17a} suggests that a change in abundance is also not likely to be an outcome of an enhanced shock chemistry in the inner regions, as shock-tracing molecules like HNCO and SiO still show elevated emission compared to the total H$_2$ column at $l>1\degr$. So, while we cannot rule out that a change in the relative \cyano\, abundance of the components leads to a larger (10-9)/(3-2) ratio in gas with $|l| < 1\degr$, there is also no obvious reason to expect that the abundance of the high- or low-excitation component would change significantly at this radius. 

\subsection{The Degeneracy Between Density and Temperature}

The radiative transfer modeling of \cyano\, excitation in Galactic center clouds yields no constraint on the gas temperature. This leads to a range of best-fitting solutions in which a solution for high densities and low temperatures appears equally likely as a solution for low densities and high temperatures for the range of lines modeled here. Due to this degeneracy between temperature and density, we assess three possible scenarios for the combination of temperature and density in each component. The impact of the temperature assumption for each of these scenarios is illustrated in Figure \ref{Fig6}, and the ranges of temperatures and densities for each scenario are reported in Table \ref{Scenarios}. 

\subsubsection{Scenario 1: The High-Excitation Component is Cold.}

From our RADEX fits, densities up to $10^7$ \cmmm\, (the upper bound on the density considered by our modeling) are allowed for the high-excitation component. Such high densities can only occur if the temperature of this component is quite low, $\lesssim$ 50 K. However, \am\, temperature studies of CMZ gas indicate that a significant fraction (50-75\%) of the CMZ gas traced by this molecule is at temperatures greater than 50 K \citep{Huttem93b,Krieger17}, much larger than the 15\% of gas that we associate with the high-excitation component. Thus, to match the \am\, observations we would have to assume that part of the low-excitation component has T$<$50 K and part has T$>$ 50 K. Further, at the position we observe in GCM0.25+0.01 the `low' kinetic temperature measured with \am\, is 60 K, which would rule out gas with densities $>10^6$ \cmmm, at least in the southern part of this cloud. 

\form\, studies have the opposite problem: here, all of the dense CMZ gas is consistent with having temperatures $>$ 65 K, with no possibility of a substantial quantity of gas existing in a colder component \citep{Ao13,Ginsburg16}.  Although cold, dense gas is not seen with \form, modeling of thermal gas properties in the CMZ environment does predict that gas at densities approaching $10^7$ \cmmm\, should, in the absence of internal heating sources like embedded protostars, begin to thermalize with dust temperatures which are $\sim$20-30 K in the CMZ \citep{Clark13,Molinari11}. Thus, while current observational evidence does not favor this scenario, it is theoretically plausible. Qualitatively, this scenario would be consistent with both cosmic ray and PDR heating, in which UV radiation and cosmic rays can more effectively penetrate and heat the low-density gas. 

\subsubsection{Scenario 2: The High-excitation Component is Hot}
In the reverse of the previous scenario, here the high-excitation/ high-density component would be hot, while the low-excitation/low-density component would be cold. This is the scenario that best matches the two-temperature model of CMZ gas from \am\, observations: the proportion of gas we find in the high/low-excitation components (15\%/85\%) is roughly consistent with that found by \cite{Krieger17}, with 20-50\% of the gas at 25-50 K, versus 50-80\% of the gas at 60-100 K.  In this scenario, the density of the low-excitation gas would be between $1-5\times10^3$ \cmmm, and the density of the high-excitation gas would be between $3\times10^4-3\times10^5$ \cmmm. Note that we expect the `cold' gas component should still be much warmer than the dust temperature, as densities $<10^4$ \cmmm\, are insufficient to thermalize the gas with the dust, and thus we would favor the `cold' gas being closer to 50 K than 25 K. Qualitatively, this scenario would be consistent with shocks as a heating mechanism, with shocks both compressing the gas to higher density, and heating it. Scenario 2 could also be consistent with some fraction of the dense gas having an internal heating source like embedded protostars. 

\subsubsection{Scenario 3: All the Gas is Hot (or Not)} 
In this scenario, the discrepancy between \am\, and \form\, temperatures is resolved by assuming that the \form\, temperatures are correct, and the \am\, temperatures are systematically low, possibly due to unaccounted-for population of the non-metastable \am\, transitions which are detected in GCM-0.02-0.07 and GCM-0.13-0.08 \citep{Mills13a}. The low-density component would have densities between $10^3$ and $10^4$ \cmmm\, and the high-density component would have densities between 1-3$\times10^5$ \cmmm.These values would however not be fully representative of typical CMZ gas, as the three clouds observed here are somewhat warmer (T$_{H2CO}$ = 90-140 K)  than generally observed in the CMZ (T$_{H2CO}$ = 65 K). The relatively low densities and high temperatures would be consistent with expectations that the gas should not be well thermalized with the dust at densities $\lesssim10^5$ \cmmm\, \citep{Clark13}. We do not consider a scenario in which all of the gas is cold, as this is ruled out based on consistent findings of substantial columns of hot gas from both \form\, and \am\, \citep{Ao13,Ginsburg16,Krieger17,Mills13a}. 

It is also possible that the gas density and temperature do not neatly co-vary, e.g., that there is both hot and cold high-density gas as well as hot and cold low-density gas. This might be expected if both cosmic rays and shocks/turbulent dissipation contribute to heating CMZ gas, and the relative strengths of these heating mechanisms vary across the CMZ. It could also occur if e.g., some of the highest density gas thermalizes with the dust and cools, while other clumps of high-density gas are internally heated by protostars, or if some of the low-density gas experiences additional PDR heating due to its proximity to \hii\, regions in the CMZ. 

\subsubsection{Distinguishing Between these Scenarios}
While we favor Scenario 2, the current data do not allow us to robustly confirm this model. Conclusively determining the temperature of the \cyano\,-emitting gas would make significant progress not just in constraining the density structure of CMZ gas, but also in distinguishing between models for heating this gas. One of the best tools for making progress on this front will be interferometric observations of \am\, and \cyano\, with ALMA and the VLA, which should be able to isolate the clumpy gas. With both high spatial and spectral resolution, it will be easier to argue that \am\, and \cyano\, are tracing the same gas, and it should be possible to well-constrain the properties of the clumpy, high-excitation component. 

\subsection{Comparison with Other Density Estimates}

Prior to this work, there have been a number of studies conducting radiative transfer modeling of non-LTE gas conditions to match observed line intensities across the CMZ to the temperatures, volume densities, and column densities responsible for their excitation. Many of these studies have been conducted with the CO molecule. \cite{Dahmen98} observe multiple isotopologues of CO 1-0 and fit the observed intensities and opacities to models with gas densities of $10^{3.0}$ \cmmm\, and temperatures of 50 K, which would apply to the bulk (most likely $\sim$2/3) of the observed CMZ gas. \cite{Nagai07} observe CO 3-2 and 1-0 and fit the observed intensities to models with gas densities of $10^{3.5-4.0}$ \cmmm, though they note they are unable to find solutions to some regions that are affected by self absorption. For comparison, \cite{Martin04} performed excitation analyses of CO 7-6 and 4-3, which have higher excitation energies than the lines probed by \cite{Nagai07} or \cite{Jones13} and find densities in cloud interiors up to $10^{4.5}$ \cmmm, which is the upper limit to which their analysis code is sensitive. Typical CMZ gas densities in excess of $10^4$ \cmmm\, have been supported by excitation analyses of H$_2$CO emission \citep{Gusten83,Zylka92} and dense gas tracers like $^{13}$CS, HCN, and \cyano\, \citep{Paglione98,Jones12,Jones13}. 

CO has also been used to infer the presence of much lower density gas. \cite{Dahmen98} estimate a third of the gas in the CMZ (and possibly up to two thirds) is in a warm and tenuous component with densities of $10^{2.0}$ \cmmm\, and temperatures of 150 K. This component is also traced by H$_3^+$ \citep{Oka05} in the near-infrared and numerous hydrides observed with Herschel \citep{Geballe10,Schilke10,Lis10,Lis10b,Sonnen13,Menten11,Monje11,Lis12}. Most of these species are observed in absorption toward multiple lines of sight against strong infrared and submillimeter continuum sources, especially toward Sgr B2 and Sgr A. The H$_3^+$ observations have characterized this gas component as warm (T$\sim$ 250-350 K), diffuse (n$\sim 10-100$ \cmmm) and pervasive-- the inferred sizes of the absorbing clouds are several tens of parsecs, and are suggested to constitute a substantial fraction of the volume filling factor in the CMZ \citep{Oka05,Goto08,Goto11}. However, as this component has thus far primarily been detected only in absorption toward pencil-beam lines of sight and in emission at extremely low spatial resolution from CO and CH \citep[e.g., 9$'$][]{Dahmen98,Magnani06,Riquelme18}, it is not clear whether it fills the entire CMZ, or if it simply comprises the more extended envelopes of the known clouds, the bulk of which lie in a $\sim$ 100 pc ring-like structure \citep{Molinari11,Kruijssen15}. 

Locally higher densities have previously been inferred from analyses of individual clouds. Observations of multiple transitions of CS in several clouds (M-0.02-0.07, and the Sickle cloud, which abuts the Quintuplet star cluster) indicate that the highest gas densities in these clouds can range from a few $10^5$ up to a few $10^6$ \cmmm\, \citep{Serabyn91,Serabyn92}. Similar excitation analyses of multiple transitions of HC$_3$N show that even denser gas can be found in the Sgr B2 cloud, whose mean density is measured to be $10^5$ cm$^{-3}$ and the core of which has densities in excess of 10$^7$ cm$^{-3}$ \citep{Morris76,Lis91}. Gas in the circumnuclear disk $\sim$1-2 parsecs from the supermassive black hole Sgr A* is also measured to be extremely dense, with gas densities from CO and HCN reaching values up to $10^5-10^6$ \cmmm \citep{RT12,Mills13b,Smith14}.

Density estimates have also been made for two of the three clouds we focus on here. Observations of \cyano\, 1-0 through 25-24 in GCM-0.02-0.07 were fit with two density components, with the lower-density component being several times 10$^3$ cm$^{-3}$  and the high-density component being a few 10$^5$ cm$^{-3}$ \citep{Walmsley86}. In GCM0.25+0.01, densities are estimated to be comparable to the highest densities in GCM-0.02-0.07, lying between $\sim8\times10^4$ \cmmm\, \citep{Longmore12} to a few times $10^5$ \cmmm\, \citep{Kauffmann13}, though these estimates are based on a combination of column density and geometric arguments, and no excitation analysis of the density has been published. Notably, all of the global gas density studies (and the majority of density studies for individual clouds) have assumed a single density component. 
	
By constraining two density components in the CMZ gas, we better quantify the distribution of CMZ gas densities that have typically been approximated as n$\sim10^4$ \cmmm\, by fitting just a single density component and yielding an indeterminately-weighted average of the true distribution of gas densities. We argue that, with $\sim$ 85\% of the \cyano-detected gas having densities of $\sim3\times10^3$ \cmmm, we should instead think of the 'typical' gas density in the Galactic center clouds as being significantly  less than $10^4$ \cmmm. In fact, by focusing on the properties of gas near the center of three of the most massive, dense clouds in the CMZ, these results may in fact be biased toward measuring higher densities than are typically present in the CMZ. Future measurements of \cyano\, lines away from the dense cloud centers would better show how representative these measured densities are.   Finally, while we do not appear to directly probe the $n\sim100$ \cmmm\, gas detected in some prior studies, our observations of densities from $\sim10^3-10^5$ \cmmm\, suggests that there may be a continuum of gas densities that connects the observed clouds to this tenuous medium (perhaps consistent with the structure of cloud envelopes) rather than sharply-defined `dense' clouds embedded in a pervasive, uniformly-diffuse medium. 

\subsection{Dense Gas and Star Formation in the CMZ}

\begin{figure}[tbh]
\includegraphics[scale=0.36]{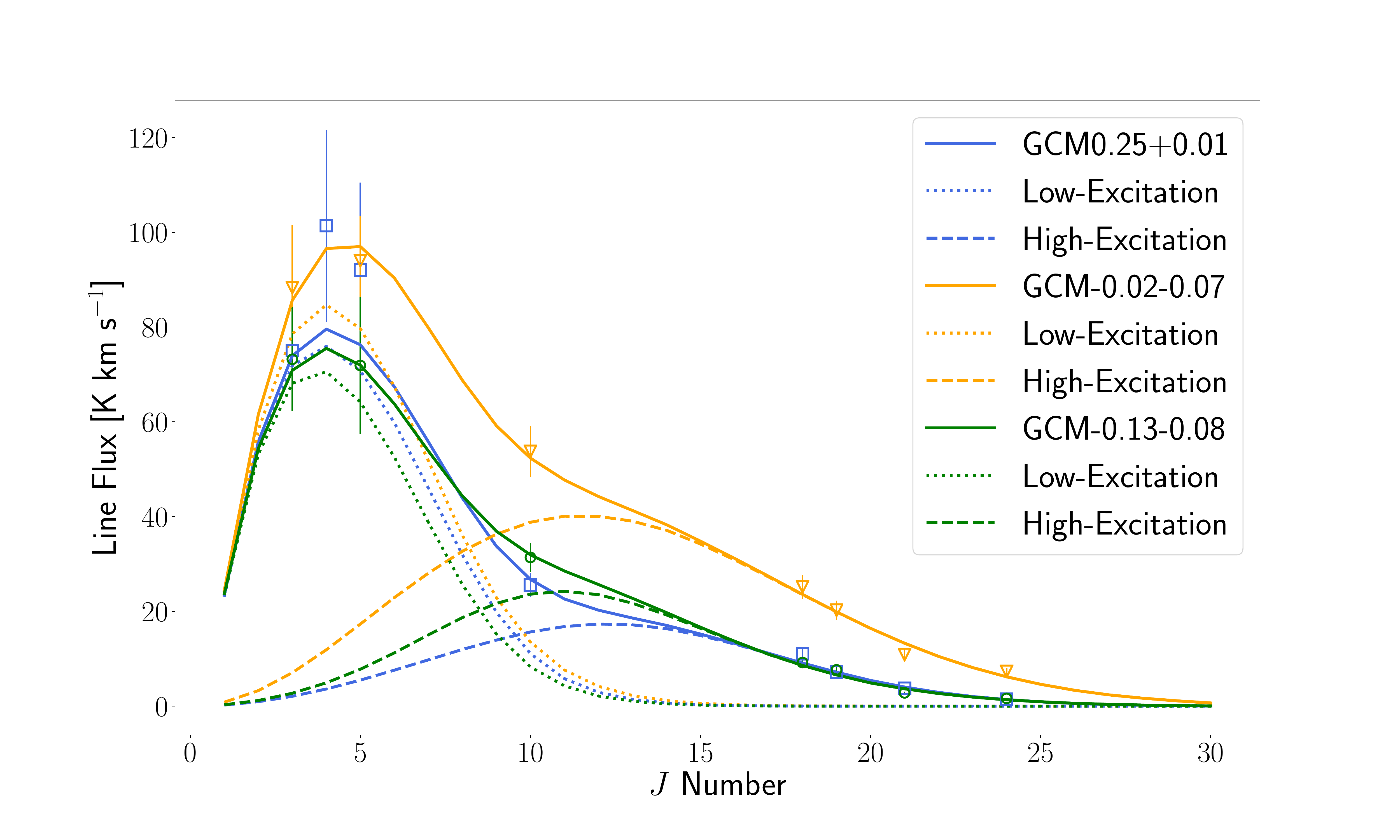}
\caption{A comparison of the best fits from RADEX (solid line) to the total line flux (individual data points) of the observed \cyano\, lines. Fits are shown for each of the three clouds: GCM 0.25+0.01, GCM-0.02-0.07, and GCM-0.13-0.08. }
\label{Fig8}
\end{figure}

Having analyzed the three clouds for which we observe the largest number of \cyano\, lines, we return now to the ratio maps of \cyano\, lines for the entire CMZ. Looking at the examples of best-fit models to the observed line intensities (Figure \ref{Fig8}), we see that the ratio of the (10-9) line to the (3-2) line should be sensitive to the presence of gas at densities above $10^4$ \cmmm, as line intensities from a $\sim10^5$ \cmmm\, density component peak around $J$=12, as can be seen in Figure \ref{Fig8}. We then interpret regions of weaker (10,9) emission as having a lower fraction of the $\gtrsim 10^4$ \cmmm\, density gas that is detected in all three of the clouds we observe.  

As noted earlier, we see a transition in the (10-9)/(3-2) ratio in Figure \ref{Fig4}, in which the (10-9) emission is relatively much weaker at $l > 1\degr$. We interpret this as there being a lower (or potentially negligible) fraction of $\gtrsim10^4$ \cmmm\, density gas outside of a radius of $\sim$ 140 pc. This is consistent with the \cite{KK17} model that predicts the gas density should sharply transition from values  $\lesssim$ 10 \cmmm\, to having an average density $\sim10^4$ \cmmm\, at this Galactocentric radius (see, e.g., their Figure 13). This is consistent with their model, which says that this  $\sim$ 100 pc radius is a minimum in the shear that transports gas inward, and is a location where gas builds up and undergoes starbursts. With the existing maps, we can only test for this signature on the positive-longitude side of the CMZ; however the \cite{KK17} model would predict that the same density transition occurs at negative longitudes beyond Sgr C as well. 

If the high-excitation gas is much hotter than 50 K (as we argue is the more likely scenario), the three observed clouds would all have peak gas densities less than $4\times10^5$\cmmm, much lower than found for the Sgr B2 cloud  \citep[$\sim10^7$ \cmmm;][]{Lis91}. If however the dense gas is at the lowest temperatures measured at these positions using \am, this could allow for gas densities in at least two of the observed clouds (GCM-0.02-0.07, and GCM-0.13-0.08) to reach values $>10^6$ \cmmm, consistent with the volume-density threshold for star formation of $\sim10^6$ \cmmm\, that is predicted by the \cite{KK17} model. Although GCM0.25+0.01, GCM-0.02-0.07, and GCM-0.13-0.08 are not actively forming stars at the same rate as Sgr B2, all three clouds do at least have some signature of star formation \citep[either water masers or compact \hii\, regions;][]{Lis94,Lu17,Ho85,Goss85}. To have the densities in these clouds then be consistent with the \cite{KK17} threshold requires either (1) that the high-density gas be at temperatures T$<50$ K, (2) that high densities are only reached in a small fraction of the gas on spatial scales much smaller than those averaged together in this study, or (3) that the densities we measure toward these three positions are not representative of the (higher) densities found elsewhere in the cloud. In the future, stronger tests of the \cite{KK17} threshold for star formation can be made both by measuring gas densities on smaller spatial scales across entire clouds, and measuring densities toward additional CMZ gas that is not associated with signatures of star formation. 

In environments outside of the Galactic center, the amount of star formation in a cloud has also been linked to the fraction of gas that exists at higher (column) densities \citep[e.g.,][]{Imara15}. ALMA observations of GCM0.25+0.01 have already shown evidence for an excess of high column density gas, deviating from a log-normal column density probability distribution function, which is associated with a star-forming core in this cloud \citep{Rathborne15}. We might expect to find a similar excess of high volume-density gas originating from star-forming structures in which self-gravity has overcome the turbulent pressure. Currently, the precision of our observations is insufficient to measure variations in the amount of high-density gas in the three observed clouds. However, we would predict a lower fraction of high-density (n$>10^5$ \cmmm) gas in clouds with little to no star formation like GCM0.25+0.01, compared to GCM-0.02-0.07, GCM-0.13-0.08, and ultimately much less than in Sgr B2. Observations of a larger number of lines (particularly including the isotopologues of all transitions to constrain the opacity in each component) should enable these measurements in the future. 

\section{Conclusion}
We have studied multiple transitions of \cyano\, in the central R$\sim$ 300 pc or CMZ of our Galaxy. We analyzed both maps of \cyano\, (3-2), (5-4) and (10-9) across the full CMZ as well as pointed observations of higher-$J$ lines toward the  GCM0.25+0.01,  GCM-0.02-0.07, and GCM-0.13-0.08 clouds. By conducting radiative transfer modeling of these lines using the RADEX code, we have arrived at the following results:

\begin{enumerate}

\item In all three clouds we find two density components, a low-excitation, low-density component with n $< 10^4$ \cmmm, and a high-excitation, high-density component with n $> 3\times10^4$ \cmmm.

\item If we adopt the measured ammonia temperatures of \cite{Krieger17} for these three clouds, assuming that the low-density gas has temperatures T = 25 - 60 K and the high-density gas has temperatures 60 -100 K,  we can further constrain the low-density component to be between $10^3$ and $10^4$ \cmmm\, and the high density component to be between $8\times10^4$ and $4\times10^5$ \cmmm. 

\item Comparing the relative columns of these two components, we find that all three clouds are consistent with having $\sim$15\% of the gas detected with \cyano\, in the high-density component. 

\item Across the entire CMZ, we find that the ratio of the \cyano\, (10-9) to (3-2) line increases sharply at R$\lesssim$ 140 pc. Based on the three clouds we modeled, we interpret this as indicating that the fraction of dense (n$>10^4$ \cmmm) gas increases inward of this radius, consistent with the predictions of the model of \cite{KK17}. 

\end{enumerate}

\section{Acknowledgements}
We thank the anonymous referee for their useful comments which improved the presentation of results in this paper. We are grateful to Cara Battersby for sharing her column density map of the CMZ in advance of publication. LW thanks Mark Morris and UCLA for support in the early parts of this project.

\bibliographystyle{hapj}
\bibliography{hc3n}

\section{Figures and Tables}
\clearpage

\clearpage

\clearpage

\clearpage

\clearpage

\clearpage

\clearpage

\begin{table}[ht]
\caption{Observed Transitions of \cyano\,} 
\centering
\begin{tabular}{cccccc}
\\[0.5ex]
\hline\hline
 & & & & &\\
 {\bf Transition} & {\bf Frequency} & {\bf E$_{\mathrm{upper}}$/k}   & {\bf n$_{\mathrm{crit}}$ at 50 K}\footnotesize{$^{\mathrm{a}}$}  & {\bf Telescope} & {\bf Beam FWHM} \\ [0.5ex]
				 		& {(GHz)} 	& {(K)} 		&$\times 10^{4}$ cm$^{-3}$ & & $''$\\
\hline
{\bfseries H$^{13}$CCCN}  & & & & & \\
\hline
  	  			 (3-2)  	&   26.450591 	&     2.5	&			& GBT 		& 28$''$ \\
				 (5-4)    	&   44.084162 	&     6.3 	&   		 	& GBT 		& 17$''$ \\
\hline
{\bfseries HC$_3$N}  & & & & & \\
\hline
  	  			 (3-2)  	&   27.29429 	&     2.6	& 0.18 			& GBT 		& 28$''$ \\
				                 &                       &              &                               & Mopra            & 120$''$\\
				 (4-3)    	&   36.39232 	&     4.2 	& 0.47 		 	& GBT 		& 21$''$ \\
				 (5-4)    	&   45.49031 	&     6.5 	& 0.96 		 	& GBT 		& 17$''$ \\
				                 &                       &              &                               & Mopra            & 65$''$\\
				 (10-9)    &   90.97902	&   24.0	& 8.2 		 	& Mopra 		& 40$''$ \\
				 (18-17)  & 163.75339   	&   74.7	& 190			& APEX 		& 38$''$ \\
				 (19-18)  & 172.84930   	&   83.0	& 220			& APEX 		& 38$''$ \\				
				 (21-20)  & 191.04030  	&  100.9	& 280			& APEX 		& 38$''$ \\				
				 (24-23)  & 218.32472	& 131.0	& $\sim$590		& APEX 		& 29$''$ \\
				& & & & &    \\
\hline
\end{tabular}
\label{HC3N}
\end{table}
\noindent\footnotesize{$^{\mathrm{a}}$Calculated using the collisional coefficients of \cite{Faure16}}
\clearpage

\begin{table}
\caption{Measured Line Parameters}
\centering
\begin{tabular}{l l l l l l l l}
\hline\hline
Source & Species & Transition & Resolution  & v$_{\mathrm{cen}}$ & v$_{\mathrm{fwhm}}$ & Peak T$_{\mathrm{MB}}$ & $\int T_{\mathrm{MB}} dv$ \\
       &         &            &             & (km s$^{-1}$)      & (km s$^{-1}$)       & (K)                  &  (km s$^{-1}$)     \\
\hline
{\bfseries GCM0.25-0.01 (a)} & H$^{13}$CCCN &  (3-2) & 28 & 33.7 $\pm$ 0.6 &  20.1 $\pm$ 1.5 &  0.14 $\pm$ 0.01 & 3.0 $\pm$ 0.2\\
                           &         &  (5-4)     & 17 & 33.4 $\pm$ 0.3 &  17.2 $\pm$ 0.8 &  0.25 $\pm$ 0.01 &   4.6 $\pm$ 0.1\\
 & HC$_3$N                  &   (3-2)    & 28 & 34.3 $\pm$ 0.1 &  18.4 $\pm$ 0.1 &  3.65 $\pm$ 0.01 & 71.3 $\pm$ 0.1\\
                           &         &  (4-3)     & 21 & 34.8 $\pm$ 0.1 &  18.7 $\pm$ 0.1 &  4.80 $\pm$ 0.02 & 95.3 $\pm$ 0.1\\
                           &         &  (5-4)     & 17 & 34.0 $\pm$ 0.1 &  19.0 $\pm$ 0.1 &  4.52 $\pm$ 0.01 & 91.5 $\pm$ 0.1\\
                           &         &  (10-9)   & 40 & 36.5 $\pm$ 0.1 &  16.7 $\pm$ 0.2 &  1.39 $\pm$ 0.01 & 24.7 $\pm$ 0.1\\
                           &         &  (18-17) & 40 & 37.0 $\pm$ 0.1 &  14.5 $\pm$ 0.1 &  0.65 $\pm$ 0.01 & 10.1 $\pm$ 0.1\\
                           &         &  (19-18) & 40 & 37.6 $\pm$ 0.1 &  14.6 $\pm$ 0.3 &  0.42 $\pm$ 0.01 &  6.5 $\pm$ 0.1\\
                           &         &  (21-20) & 40 & 36.6 $\pm$ 0.2 &  15.2 $\pm$ 0.6 &  0.22 $\pm$ 0.01 &  3.6 $\pm$ 0.1\\
                           &         &  (24-23) & 40 & 35.7 $\pm$ 0.7 &  15.1 $\pm$ 1.7 &  0.09 $\pm$ 0.01 &  1.5 $\pm$ 0.2\\
\hline
 {\bfseries GCM0.25-0.01 (b)}  & HC$_3$N &   (3-2) & 28 & 20.0 &  23.0 $\pm$ 0.8 &  0.33 $\pm$ 0.01 &  8.1 $\pm$ 0.4\\
                           &         &  (4-3)     & 21 & 20.0                  &  18.9 $\pm$ 1.3 &  0.38 $\pm$ 0.03 &  7.7 $\pm$ 0.8\\
                           &         &  (5-4)     & 17 & 20.0                  &  14.5 $\pm$ 1.1 &  0.13 $\pm$ 0.01 &  2.0 $\pm$ 0.3\\
                           &         &  (10-9)   & 40 & 20.0                  &  12.1 $\pm$ 2.0 &  0.14 $\pm$ 0.02 &  1.8 $\pm$ 0.4\\
                           &         &  (18-17) & 40 & 20.0                  &  10.2 $\pm$ 0.7 &  0.11 $\pm$ 0.01 &  1.2 $\pm$ 0.1\\
                           &         &  (19-18) & 40 & 20.0                  &  12.0 $\pm$ 1.6 &  0.06 $\pm$ 0.01 &  0.8 $\pm$ 0.1\\
                           &         &  (21-20) & 40 & 20.0                  &  11.6 $\pm$ 5.8 &  0.02 $\pm$ 0.01 &  0.2 $\pm$ 0.2\\
 \hline
{\bfseries GCM-0.13-0.08 (a)} & H$^{13}$CCCN &  (3-2) & 28 & 7.0 $\pm$ 0.1 &  19.0 $\pm$ 2.9 &  0.07 $\pm$ 0.01 & 1.5 $\pm$ 0.3\\
                           &         &  (5-4)     & 17 &   1.5 $\pm$ 1.3 &  13.3 $\pm$ 2.9 &  0.08 $\pm$ 0.01 &   1.2 $\pm$ 0.3\\
& HC$_3$N                   &  (3-2)     & 28 &   7.0 $\pm$ 0.1 &  22.0 $\pm$ 0.4 &  2.29 $\pm$ 0.04 & 53.7 $\pm$ 0.2\\
                           &         &  (5-4)     & 17 &   5.7 $\pm$ 0.1 &  20.6 $\pm$ 0.1 &  2.50 $\pm$ 0.01 & 54.6 $\pm$ 0.1\\
                           &         &  (10-9)   & 40 &   7.0 $\pm$ 0.1 &  15.9 $\pm$ 0.2 &  1.78 $\pm$ 0.02 & 30.1 $\pm$ 0.1\\
                           &         &  (18-17) & 40 &   6.7 $\pm$ 0.1 &  16.8 $\pm$ 0.3 &  0.52 $\pm$ 0.01 &   9.3 $\pm$ 0.1\\
                           &         &  (19-18) & 40 &   7.4 $\pm$ 0.1 &  18.0 $\pm$ 0.3 &  0.40 $\pm$ 0.01 &   7.6 $\pm$ 0.1\\
                           &         &  (21-20) & 40 &   6.4 $\pm$ 0.4 &  16.3 $\pm$ 1.0 &  0.17 $\pm$ 0.01 &   2.9 $\pm$ 0.1\\
                           &         &  (24-23) & 40 &   6.0 $\pm$ 0.6 &  13.0 $\pm$ 1.5 &  0.11 $\pm$ 0.01 &   1.5 $\pm$ 0.2\\
\hline
{\bfseries GCM-0.13-0.08 (b)} & H$^{13}$CCCN &  (3-2) & 28 & 12.5 $\pm$ 0.6 &  5.4 $\pm$ 0.9 &  0.13 $\pm$ 0.02 &  0.7 $\pm$ 0.2\\
 &                  &  (5-4) & 17 &  11.7 $\pm$ 0.2 &  6.3 $\pm$ 0.5 &  0.24 $\pm$ 0.02 &  1.6 $\pm$ 0.2\\
 & HC$_3$N &  (3-2) & 28 & 12.7 $\pm$ 0.1 &  6.0 $\pm$ 0.1 &  2.87 $\pm$ 0.07 &  18.4 $\pm$ 0.4\\
 &                  &  (5-4) & 17 &  12.7 $\pm$ 0.1 &  5.7 $\pm$ 0.1 &  2.57 $\pm$ 0.01 &  15.5 $\pm$ 0.1\\
 \hline
{\bfseries GCM-0.02-0.07} & H$^{13}$CCCN &  (3-2) & 28 & 50.2 $\pm$ 0.8 &  30.0 $\pm$ 0.1 &  0.13 $\pm$ 0.01 &  4.1 $\pm$ 0.2\\
 &                  &  (5-4) & 40 & 51.2 $\pm$ 0.5 &  23.8 $\pm$ 1.2 &  0.18 $\pm$ 0.01 &  4.7 $\pm$ 0.3\\
 & HC$_3$N &  (3-2) & 28 & 48.9 $\pm$ 0.1 &  24.6 $\pm$ 0.1 &  3.31 $\pm$ 0.01 &  86.6 $\pm$ 0.2\\
 &                  &  (5-4) & 40 & 50.6 $\pm$ 0.1 &  23.4 $\pm$ 0.1 &  3.75 $\pm$ 0.02 &  93.7 $\pm$ 0.8\\
 &                  &  (10-9) & 40 & 47.1 $\pm$ 0.1 &  24.3 $\pm$ 0.2 &  2.04 $\pm$ 0.01 &  52.8 $\pm$ 0.5\\
 &                  &  (18-17) & 40 & 47.1 $\pm$ 0.1 &  21.1 $\pm$ 0.1 &  1.12 $\pm$ 0.01 &  25.0 $\pm$ 0.2\\
 &                  &  (19-18) & 40 & 47.2 $\pm$ 0.1 &  20.8 $\pm$ 0.2 &  0.90 $\pm$ 0.01 &  20.0 $\pm$ 0.2\\
 &                  &  (21-20) & 40 & 47.1 $\pm$ 0.1 &  20.9 $\pm$ 0.4 &  0.47 $\pm$ 0.01 &  10.4 $\pm$ 0.2\\
 &                  &  (24-23) & 40 & 44.4 $\pm$ 0.3 &  21.4 $\pm$ 0.7 &  0.31 $\pm$ 0.01 &  7.0 $\pm$ 0.3\\
\hline\hline
\end{tabular}
\label{line_params}
\end{table}
\clearpage

 \begin{table}[ht]
\caption{Adopted and Modeled Source Properties} 
\centering
\begin{tabular}{lc|ccc}
\\[0.5ex]
\hline\hline
 & & & & \\
& {\bf Excitation}& {\bf GCM0.25+0.01} & {\bf GCM-0.02-0.07} & {\bf GCM-0.13-0.08} \\ [0.5ex]
 & {\bf Component}& & &  \\ [0.5ex]
\hline
 {\bf log [N$_{\mathrm{HC}_3\mathrm{N}}$(cm$^{-2}$)]} & low & $14.9^{+0.2}_{-0.5}$ & $14.9^{+0.2}_{-0.5}$ & $14.9^{+0.1}_{-0.5}$ \\
 {\bf log [N$_{\mathrm{HC}_3\mathrm{N}}$(cm$^{-2}$)]}  & high & $13.8^{+0.3}_{-0.4}$ & $14.2^{+0.1}_{-0.5}$ & $13.9^{+0.1}_{-0.5}$\\
 {\bf N$_{\mathrm{HC}_3\mathrm{N (high)}}$ / N$_{\mathrm{HC}_3\mathrm{N (low)}}$} & & 0.08$\pm$0.06 & 0.20$\pm$0.15 & 0.10$\pm$0.09 \\
 {\bf log [N$_{\mathrm{H}2}$(cm$^{-2}$)]\footnotesize{$^{\mathrm{a}}$}}  & both & $22.40^{+0.07}_{-0.10}$ & $22.28^{+0.04}_{-0.05}$ & $22.67^{+0.08}_{-0.09}$  \\
 {\bf log [HC$_3$N/H$_2$]}  & both &  $-8.5^{+0.2}_{-0.4}$ &  $-8.3^{+0.1}_{-0.3}$ &  $-8.8^{+0.2}_{-0.4}$ \\
 
 \hline
 {\bf T$_\mathrm{kin,NH3}$ (K)\footnotesize{$^{\mathrm{b}}$} }  & low & 64 $\pm$ 18  & 41 $\pm$ 8  & 26 $\pm$ 3 \\
  {\bf T$_\mathrm{kin,NH3}$ (K)\footnotesize{$^{\mathrm{c}}$} } &  high & 69 $\pm$ 4  & 102 $\pm$ 7  & 54 $\pm$ 6 \\
 {\bf T$_\mathrm{kin,H2CO}$ (K)\footnotesize{$^{\mathrm{d}}$} }  & both &140 $\pm$ 30  & 101 $\pm$ 13  & 91 $\pm$ 16 \\
 
   \hline
 {\bf log [n$_{\mathrm{H}2}$(cm$^{-3}$)]} & low  &  3.5$^{+0.5}_{-0.4}$ & 3.4$^{+0.7}_{-0.2}$  & 3.4$^{+0.7}_{-0.3}$ \\
 {\bf log [n$_{\mathrm{H}2}$(cm$^{-3}$)]} & high & $>$4.6 & $>$4.7 & $>$4.5 \\ 
 {\bf log [n$_{\mathrm{H}2}$(cm$^{-3}$)]\footnotesize{$^{\mathrm{e}}$} } & low & 3.3$^{+0.3}_{-0.1}$ & 3.4$^{+0.3}_{-0.2}$  & 3.3$^{+0.5}_{-0.2}$ \\
 {\bf log [n$_{\mathrm{H}2}$(cm$^{-3}$)]\footnotesize{$^{\mathrm{f}}$}} & high & 5.2$^{+0.3}_{-0.2}$ & $>$5.6 & $>$5.5 \\
 {\bf log [n$_{\mathrm{H}2}$(cm$^{-3}$)]\footnotesize{$^{\mathrm{g}}$}} & low & 3.3$^{+0.3}_{-0.1}$ & 3.4$^{+0.6}_{-0.1}$  & 3.4$^{+0.6}_{-0.2}$ \\
 {\bf log [n$_{\mathrm{H}2}$(cm$^{-3}$)]\footnotesize{$^{\mathrm{h}}$}} & high & 5.2$^{+0.3}_{-0.3}$ & 5.2$^{+0.5}_{-0.3}$  & 5.2$^{+0.5}_{-0.3}$\\
 {\bf log [n$_{\mathrm{H}2}$(cm$^{-3}$)]\footnotesize{$^{\mathrm{i}}$}} & low & 3.2$^{+0.3}_{-0.2}$ & 3.4$^{+0.3}_{-0.2}$  & 3.3$^{+0.4}_{-0.2}$ \\
 {\bf log [n$_{\mathrm{H}2}$(cm$^{-3}$)]\footnotesize{$^{\mathrm{i}}$}} & high & 4.9$^{+0.1}_{-0.2}$ & 5.2$^{+0.4}_{-0.3}$  & 4.8$^{+0.5}_{-0.1}$\\
 
\hline
\end{tabular}
\label{Scenarios}
\end{table}
\noindent\footnotesize{$^{\mathrm{a}}$From Battersby et al. (in prep.)  derived using Herschel data using the method of \cite{Battersby11}.\\
\noindent\footnotesize{$^{\mathrm{b}}$From \cite{Krieger17}, derived from the (1,1)-(2,2) lines of NH$_3$}.\\
\noindent\footnotesize{$^{\mathrm{c}}$From \cite{Krieger17}, derived from the mean of the temperatures between the (2,2)-(4,4),  (3,3)-(6,6), and (4,4)-(5,5) lines of NH$_3$}.\\
\noindent\footnotesize{$^{\mathrm{d}}$From \cite{Ginsburg16} derived from the 3$_{21}$-2$_{20}$ and 3$_{03}$-2$_{02}$ lines of H$_2$CO}.\\
\noindent\footnotesize{$^{\mathrm{e}}$Determined by restricting temperature to the mean value measured by \cite{Krieger17} using the $J>2$ \am\, lines (Scenario 1).\\
\noindent\footnotesize{$^{\mathrm{f}}$Determined by restricting temperature to the value measured by \cite{Krieger17} using the (1,1)-(2,2) \am\, lines (Scenario 1).\\
\noindent\footnotesize{$^{\mathrm{g}}$Determined by restricting temperature to the value measured by \cite{Krieger17} using the (1,1)-(2,2) \am\, lines (Scenario 2).\\
\noindent\footnotesize{$^{\mathrm{h}}$Determined by restricting temperature to the mean value measured by \cite{Krieger17} using the $J>2$ \am\, lines (Scenario 2).\\
\noindent\footnotesize{$^{\mathrm{i}}$Determined by restricting temperature to the mean value measured by \cite{Ginsburg16} (Scenario 3).\\

\clearpage

\end{document}